\RequirePackage{ifpdf}
\ifpdf 
\documentclass[pdftex]{sigma}
\else
\documentclass{sigma}
\fi

\newtheorem{Theorem}{Theorem}[section]
\newtheorem{Proposition}[Theorem]{Proposition}
\newtheorem{Corollary}[Theorem]{Corollary}
{\theoremstyle{definition}
\newtheorem{Definition}[Theorem]{Definition}

\newtheorem{Remark}[Theorem]{Remark}
\newtheorem{Example}[Theorem]{Example}
}

\usepackage[mathscr]{eucal}   						

\DeclareMathOperator{\CCl}{\CC l}		

\DeclareMathOperator{\im}{Im}
\DeclareMathOperator{\spa}{span}

\DeclareMathOperator{\Ob}{Ob}

\DeclareMathOperator{\Hom}{Hom}

\DeclareMathOperator{\dom}{Dom}
\DeclareMathOperator{\Aut}{Aut}
\DeclareMathOperator{\Sp}{Sp}

\DeclareMathOperator{\Ad}{Ad}

\newcommand{\cj}[1]{\overline{#1}}									
\newcommand{\mip}[2]{(#1\mid #2)}									
\newcommand{\ip}[2]{\langle #1\mid #2\rangle}					
\renewcommand{\iff}{\Leftrightarrow}								
\newcommand{\imp}{\Rightarrow}										

\newcommand{\A}{\mathcal{A}}\newcommand{\B}{\mathcal{B}}

\newcommand{\G}{\mathcal{G}}
\renewcommand{\H}{\mathcal{H}} 	
\newcommand{\K}{\mathcal{K}}
\newcommand{\M}{\mathcal{M}}\newcommand{\N}{\mathcal{N}}
\renewcommand{\O}{\mathcal{O}}	
\renewcommand{\P}{\mathcal{P}}	

\newcommand{\W}{\mathcal{W}}

\newcommand{\As}{{\mathscr{A}}}\newcommand{\Bs}{{\mathscr{B}}}\newcommand{\Cs}{{\mathscr{C}}}
\newcommand{\Ds}{{\mathscr{D}}}
\newcommand{\Fs}{{\mathscr{F}}}

\newcommand{\Ms}{{\mathscr{M}}}\newcommand{\Ns}{{\mathscr{N}}}\newcommand{\Os}{{\mathscr{O}}}
\newcommand{\Rs}{{\mathscr{R}}}\newcommand{\Ss}{{\mathscr{S}}}
\newcommand{\Ws}{{\mathscr{W}}}

\newcommand{\CC}{{\mathbb{C}}}

\newcommand{\MM}{{\mathbb{M}}}
\newcommand{\NN}{{\mathbb{N}}}

\newcommand{\RR}{{\mathbb{R}}}
\newcommand{\TT}{{\mathbb{T}}}

\newcommand{\ZZ}{{\mathbb{Z}}}

\newcommand{\Ag}{\mathfrak{A}}

\newcommand{\Gg}{\mathfrak{G}}

\newcommand{\Lg}{\mathfrak{L}}
\newcommand{\Mg}{\mathfrak{M}}

\DeclareFontFamily{U}{rsfs}{\skewchar\font127 }
\DeclareFontShape{U}{rsfs}{m}{n}{%
   <5> <6> rsfs5
   <7> rsfs7
   <8> <9> <10> <10.95> <12> <14.4> <17.28> <20.74> <24.88> rsfs10
}{}
\DeclareSymbolFont{rsfs}{U}{rsfs}{m}{n}
\DeclareSymbolFontAlphabet{\scr}{rsfs}

\newcommand{\Bf}{\scr{B}}
\newcommand{\Hf}{\scr{H}}

\begin{document}

\allowdisplaybreaks

\renewcommand{\thefootnote}{$\star$}

\renewcommand{\PaperNumber}{067}

\FirstPageHeading

\ShortArticleName{Modular Theory, Non-Commutative Geometry and Quantum Gravity}

\ArticleName{Modular Theory, Non-Commutative Geometry\\ and   Quantum Gravity\footnote{This paper is a
contribution to the Special Issue ``Noncommutative Spaces and Fields''. The
full collection is available at
\href{http://www.emis.de/journals/SIGMA/noncommutative.html}{http://www.emis.de/journals/SIGMA/noncommutative.html}}}

\Author{Paolo BERTOZZINI~$^\dag$, Roberto CONTI~$^\ddag$ and Wicharn LEWKEERATIYUTKUL~$^\S$}

\AuthorNameForHeading{P.~Bertozzini, R.~Conti and W.~Lewkeeratiyutkul}

\Address{$^\dag$~Department of Mathematics and Statistics,
Faculty of Science and Technology,\\
\hphantom{$^\dag$}~Thammasat University, Pathumthani 12121, Thailand}
\EmailD{\href{mailto:paolo.th@gmail.com}{paolo.th@gmail.com}}
\URLaddressD{\url{http://www.paolo-th.110mb.com}}

\Address{$^\ddag$~Dipartimento di Scienze,
Universit\`a di Chieti-Pescara ``G. D'Annunzio'',
\\
\phantom{$^\ddag$}~Viale Pindaro 42, I-65127 Pescara, Italy}
\EmailD{\href{mailto:conti@sci.unich.it}{conti@sci.unich.it}}

\Address{$^\S$~Department of Mathematics, Faculty of Science,
Chulalongkorn University,
\\
\phantom{$^\ddag$}~Bangkok 10330, Thailand}
\EmailD{\href{mailto:Wicharn.L@chula.ac.th}{Wicharn.L@chula.ac.th}}
\URLaddressD{\url{http://www.math.sc.chula.ac.th/wicharn.l}}

\ArticleDates{Received March 30, 2010, in f\/inal form July 26, 2010;  Published online August 19, 2010}

\Abstract{This paper contains the f\/irst written exposition of some ideas (announced in a~previous survey) on an approach to
quantum gravity based on Tomita--Takesaki modular theory and A.~Connes non-commutative geometry aiming at the reconstruction of spectral geometries from an operational formalism of states and categories of observables in a~covariant theory.
Care has been taken to provide a coverage of the relevant background on modular theory, its applications in
non-commutative geometry and physics and to the detailed discussion of the main foundational issues raised by the proposal.}


\Classification{46L87;			
					46L51;
					46L10;
					46M15; 			
					18F99; 			
					58B34; 			
					81R60;			
					81T05;			
					83C65}			

\tableofcontents

\renewcommand{\thefootnote}{\arabic{footnote}}
\setcounter{footnote}{0}

\section{Introduction}

This paper is an expansion of the physical implications of some ideas already sketched in the last part of the companion
survey~\cite{BCL6} and aimed at setting up an approach to non-perturbative quantum gravity based on Tomita--Takesaki modular theory.
We also provide an exhaustive discussion of background material, notably modular theory and non-commutative geometry, that could be useful to put these ideas in the right perspective and to stimulate crossing relations and further development.

In Section~\ref{sec: tt}, in order to establish our notation, we describe some well-known notions in the theory of operator algebras and some (maybe not so well-known) portions of Tomita--Takesaki modular theory. The emphasis is on potentially useful categorical and duality results that in our opinion have not been fully exploited in their applications to physics and that we expect to turn out to be useful tools in our proposal for modular algebraic quantum gravity.

A sharp transition from mathematics to physics characterizes Section~\ref{sec: mtp}, where we have tried to summarize (with some attention to actual historical developments) the current applications of Tomita--Takesaki modular theory in mathematical physics.
We hope that this material might be useful to appropriately contextualize our discussion and proposal for modular algebraic quantum gravity in the subsequent Section~\ref{sec: maqg}. Care has been taken to provide at least some essential bibliographic references.

In Section~\ref{sec: ncg}, always for notational purposes and to make the paper as self-contained as pos\-sible, we brief\/ly introduce the basic notions of metric non-commutative geometry that, following A.~Connes, are specif\/ied via axioms for spectral triples.
We also mention some alternative proposals of axiomatization for non-commutative geometries, such as Lorentzian spectral triples and especially non-commutative Riemannian geometries and phase spaces, that already show closer links with modular theory.

Section~\ref{sec: mncg} contains a discussion of the interplay between A.~Connes' non-commutative geo\-metry and Tomita--Takesaki modular theory.
Most of the material is inspired by the work of A.~Carey, A.~Rennie, J.~Phillips and F.~Sukochev, but we brief\/ly present other approaches  including some material, that was developed a few years ago by the authors, on the interplay between modular theory and spectral triples for AF $C^*$-algebras constructed by C.~Antonescu and E.~Christensen.
In a small subsection we present some bibliographic references for those few situations in which modular theory and non-commutative geometry together have already started to f\/ind connections with physics.

The main focus of this work is the f\/inal Section~\ref{sec: maqg} in which we explain the philosophical motivations for the spectral  reconstruction of non-commutative geometries from states over cate\-go\-ries of observable algebras.
A f\/irst connection between our modular spectral geometries and A.~Carey, J.~Phillips and A.~Rennie modular spectral triples is established in Theorem~\ref{th: cpr}, while some elementary steps in the direction of applications to loop quantum gravity are developed in Subsection~\ref{sec: otherqg}.
Most of the steps described here are still tentative and the reader might feel disappointed f\/inding that no solution will be of\/fered here to the deep problems of reconciliation between quantum physics and generally relativistic physics, but we hope that at least an alternative path to the physical interpretation of modular covariance might be for now suf\/f\/iciently intriguing.

The reader is warned that the paper is divided into three strictly interrelated, but essentially dif\/ferent parts, as long as their established relevance for physics is concerned.

Sections~\ref{sec: tt} and~\ref{sec: mtp} deal with a review of standard mathematical results from Tomita--Takesaki modular theory and their already well-established and absolutely unquestionable applications in mathematical physics, notably quantum statistical mechanics and algebraic quantum f\/ield theory. A few more speculative physical applications are mentioned as well, but only if they are based on the usage of standard modular theory alone.

Sections~\ref{sec: ncg} and~\ref{sec: mncg} contain a review of mostly purely mathematical material (spectral triples) in the area of A.~Connes' non-commutative geometry and explore in some detail several possible connections that have been emerging with Tomita--Takesaki modular theory. It is important to stress that, despite the perfectly sound framework, the astonishing mathematical achievements and the strong appeal of non-commutative geometry among theoretical physicists, ideas from non-commutative geometry are still lacking the same solid validation of modular theory and their basic role in physics is still somehow  questionable.

Finally, Section~\ref{sec: maqg} contains (for now) extremely speculative material on a possible marriage between modular theory and non-commutative geometry in pursue of a fundamental theory of quantum physics.
Although the proposed framework is still incomplete, to say the least,
we think that a detailed exposition of these ideas (originally formulated several years ago) is already overdue and somehow urgent in order to make connection with current developments in quantum gravity research.

\section{Tomita--Takesaki modular theory} \label{sec: tt}

\subsection{Operator algebras}

Just for the purpose of establishing notation, conventions and some terminology, we introduce here some basic notions from the theory of operator algebras\footnote{For a general overview of the theory of operator algebras we refer the reader to the reference book by B.~Blackadar~\cite{Bl} and, among the many textbooks, to the detailed treatments in R.~Kadison and J.~Ringrose~\cite{KR},
B.R.~Li~\cite{B-R}, J.~Fell and R.~Doran~\cite{FD}, M.~Takesaki~\cite{T}. Our exposition of dynamical systems and KMS-states can be found in O.~Bratteli and D.~Robinson~\cite[Sections~2.5.3, 2.7.1, 5.3]{BR}.}.

A complex unital \emph{algebra} $\As$ is a vector space over $\CC$ with an associative unital bilinear multiplication.
$\As$ is \emph{Abelian} (commutative) if $ab=ba$, for all $a,b\in \As$.

A complex algebra $\As$ is called an \emph{involutive algebra} (or $*$-algebra) if it is equipped with  an \emph{involution} i.e.~a conjugate linear map $*: \As\to \As$ such that $(a^*)^*=a$ and $(ab)^*=b^*a^*$, for all $a,b\in \As$.

An involutive complex unital algebra $\As$ is called a \emph{$C^*$-algebra} if $\As$ is a Banach space with a~norm $a\mapsto\|a\|$ such that $\|ab\|\leq\|a\|\cdot\|b\|$ and  $\|a^*a\|=\|a\|^2$, for all $a,b\in\As$.

Notable examples are the algebras of continuous complex valued functions $C(X;\CC)$ on a~compact topological space $X$ with the ``sup norm'' $\|f\|:=\sup_{p\in X}|f(p)|$, for all $f\in C(X;\CC)$, and the algebras of linear bounded operators $\B(\H)$ on a given Hilbert space $\H$.

A \emph{von Neumann algebra} $\M\subset\B(\H)$ is a $C^*$-algebra acting on the Hilbert space $\H$ that is closed under the weak-operator topology:
$A_n \xrightarrow{n\to\infty} A\quad  \text{if\/f}\quad \ip{\xi}{A_n\eta}\xrightarrow{n\to\infty}\ip{\xi}{A\eta}$, for all $\xi,\eta\in\H$ or equivalently under the \emph{$\sigma$-weak topology}:
$A_n \xrightarrow{n\to\infty} A$ if and only if for all sequences $(\xi_k),(\zeta_k)$ in $\H$ such that
$\sum_{k=1}^{+\infty}\|\xi_k\|^2<+\infty$ and $\sum_{k=1}^{+\infty}\|\zeta_k\|^2<+\infty$ we have
$\sum_{k=1}^{+\infty}\ip{\xi_k}{A_n\zeta_k}\xrightarrow{n\to+\infty}\sum_{k=1}^{+\infty}\ip{\xi_k}{A\zeta_k}$.

The \emph{pre-dual} $\M_*$ of a von Neumann algebra $\M$ is the set of all $\sigma$-weakly continuous functionals on $\M$.
It is a Banach subspace of the dual $\M^*$.  The von Neumann algebra $\M$ is always the dual of $\M_*$.
By a theorem of S.~Sakai, a $C^*$-algebra $\As$ is isomorphic to a von Neumann algebra $\M$ if and only if it is a dual of a Banach space.

A \emph{state} $\omega$ over a unital $C^*$-algebra $\As$ is a linear function $\omega:\As\to\CC$ that is positive $\omega(x^*x)\geq 0$ for all $x\in \As$ and normalized $\omega(1_\As)=1$.

To every state $\omega$ over a unital $C^*$-algebra $\As$ we can associate its \emph{Gel'fand--Na\u\i mark--Segal representation}, i.e.~a unital $*$-homomorphism $\pi_\omega:\As\to\B(\H_\omega)$, over a Hilbert space $\H_\omega$ with a~norm-one vector
$\xi_\omega$ such that $\omega(x)=\ip{\xi_\omega}{\pi_\omega(x)\xi_\omega}$ for all $x\in \As$.

A \emph{$C^*$-dynamical system} $(\As,\alpha)$ is a $C^*$-algebra $\As$ equipped with a group homomorphism $\alpha:G\to\Aut(\As)$ that is strongly continuous i.e.~$g\mapsto\|\alpha_g(x)\|$ is a continuous map for all $x\in \As$. Similarly a \emph{von Neumann dynamical system} $(\M,\alpha)$ is a von Neumann algebra acting on the Hilbert space $\H$ equipped with a group homomorphism
$\alpha:G\to\Aut(\M)$ that is weakly continuous i.e.~$g\mapsto \ip{\xi}{\alpha_g(x)\eta}$ is continuous for all $x\in \M$ and all
$\xi,\eta\in \H$.

For a \emph{one-parameter} ($C^*$ or von Neumann) dynamical system $(\As,\alpha)$, with $\alpha:\RR\to\Aut(\As)$, an element
$x\in \As$ is \emph{$\alpha$-analytic} if there exists a holomorphic extension of the map $t\mapsto \alpha_t(x)$ to an open horizontal strip $\{z\in \CC \ | \ |\im z|<r\}$, with $r>0$, in the complex plane. The set of $\alpha$-analytic elements is always
$\alpha$-invariant (i.e.~for all $x$ analytic, $\alpha(x)$ is analytic) $*$-subalgebra of $\As$ that is norm dense in the $C^*$ case and weakly dense in the von Neumann case.

A state $\omega$ on a one-parameter $C^*$-dynamical system $(\As,\alpha)$ is a \emph{$(\alpha,\beta)$-KMS state}, for $\beta\in \RR$, if for all pairs of elements $x$, $y$ in a norm dense $\alpha$-invariant $*$-subalgebra of $\alpha$-analytic elements of $\As$ we have
$\omega(x\alpha_{i\beta}(y))=\omega(yx)$. In the case of a von Neumann dynamical system $(\M,\alpha)$, a~$(\alpha,\beta)$-KMS state must be normal and should satisfy the above property for all pairs of elements in a weakly dense $\alpha$-invariant $*$-subalgebra of $\alpha$-analytic elements of $\M$.

\subsection{Modular theory}\label{sec: m2}

The \emph{modular theory}\footnote{Among the several introductions to modular theory now available, the reader can consult
\c S.~Str\u atil\u a, L.~Zsid\'o~\cite[Chapter~10]{SZ}, \c S.~Str\u atil\u a~\cite{St}, S.~Doplicher's seminar notes~\cite[Chapter IV]{BDNR}, O.~Bratteli, D.~Robinson~\cite[Sections 2.5, 2.7.3, 5.3]{BR}, S.~Sunder~\cite{Sun},
R.~Kadison, J.~Ringrose~\cite[Chapter~9]{KR},  B.R.~Li~\cite[Chapter~8]{B-R}, M.~Takesaki~\cite[Vol.~II]{T},
B.~Blackadar~\cite[Section~III.4]{Bl}.
For introductions to the applications of modular theory to mathematical physics some of the best sources are:
R.~Haag~\cite[Chapter V]{H}, R.~Longo~\cite{L1}, H.-J.~Borchers~\cite{Bo2} and especially the recent survey papers by
S.J.~Summers~\cite{Su1}, D.~Guido~\cite{G}, F.~Lled\'o~\cite{Lle2}.}
of von Neumann algebras has been discovered by
M.~Tomita~\cite{To1,To2} in 1967 and put on solid grounds by M.~Takesaki~\cite{T1} around 1970.
It is a very deep theory that, to every von Neumann algebra $\M\subset \B(\H)$ acting on a Hilbert space $\H$, and to every vector
$\xi\in \H$ that is \emph{cyclic}, i.e.~$\cj{(\M \xi)}=\H$, and \emph{separating}, i.e.~for $A\in \M$, $A\xi=0 \imp A=0$, associates:
\begin{itemize}\itemsep=0pt
\item
a one-parameter unitary group $t \mapsto \Delta^{it}\in \B(\H)$ \item
and a conjugate-linear isometry $J:\H \to \H$ such that:{\samepage
\begin{gather*}
\Delta^{it}\M\Delta^{-it}= \M, \quad \forall\, t \in \RR,
\qquad \text{and} \qquad
J\M J = \M',
\end{gather*}where the \emph{commutant} $\M'$ of $\M$ is def\/ined by $\M'\!:=\!\{A' \in \B(\H) \, | \, [A',A]_{-}\!=0, \forall \, A \!\in\! \B(\H)\}$.}
\end{itemize}

More generally, given a von Neumann algebra $\M$ and a faithful normal state\footnote{$\omega$ is faithful if
$\omega(x)=0 \imp x=0$; it is normal if for every increasing bounded net of positive elements $x_\lambda\to x$, we have
$\omega(x_\lambda)\to\omega(x)$.}
(more generally for a faithful normal semi-f\/inite weight) $\omega$ on the algebra $\M$, the modular theory allows to create
a one-parameter group of $*$-automorphisms of the algebra $\M$,
\begin{equation*}
\sigma^{\omega}: t\mapsto \sigma^\omega_t \in \text{Aut}(\M), \qquad \text{with}\quad t \in \RR,
\end{equation*}
such that:
\begin{itemize}\itemsep=0pt
\item
in the Gel'fand--Na\u\i mark--Segal representation $\pi_\omega$ induced by the weight $\omega$, on the Hilbert space $\H_\omega$, the \emph{modular automorphism group} $\sigma^\omega$ is implemented by a unitary one-parameter group
$t \mapsto \Delta_\omega^{it}\in \B(\H_\omega)$
i.e.~we have $\pi_\omega(\sigma^\omega_t(x))=\Delta^{it}_\omega\pi_\omega(x)\Delta_\omega^{-it}$, for all $x \in \M$ and for all $t \in \RR$;
\item
there is a conjugate-linear isometry $J_\omega:\H_\omega \to \H_\omega$, whose adjoint action implements a~\emph{modular anti-isomorphism} $\gamma_\omega: \pi_\omega(\M)\to \pi_\omega(\M)'$, between $\pi_\omega(\M)$ and its commutant $\pi_\omega(\M)'$, i.e.~for all $x\in \M$, we have $\gamma_\omega(\pi_\omega(x))=J_\omega \pi_\omega(x)J_\omega$.
\end{itemize}

The operators $J_\omega$ and $\Delta_\omega$ are called respectively the \emph{modular conjugation operator} and the \emph{modular operator} induced by the state (weight) $\omega$.
We will call ``\emph{modular generator}'' the self-adjoint generator of the unitary one-parameter group
$t\mapsto \Delta_\omega^{it}$ as def\/ined by Stone's theorem i.e.~the operator
\begin{equation*}
K_\omega:= \log \Delta_\omega, \qquad \text{so that} \quad  \Delta_\omega^{it}=e^{iK_\omega t}.
\end{equation*}

The modular automorphism group $\sigma^\omega$ associated to $\omega$ is the unique one-parameter automorphism group that satisf\/ies the Kubo--Martin--Schwinger \emph{KMS-condition} with respect to the state (or more generally a normal semi-f\/inite faithful weight) $\omega$, at inverse temperature $\beta=-1$, i.e.
\begin{equation*}
\omega(\sigma^\omega_t(x))=\omega(x), \qquad \forall\, x\in \M
\end{equation*}
and for all $x,y\in \M$, there exists a function $F_{x,y}: \RR\times [0,\beta]\to \CC$ such that:
\begin{gather*}
F_{x,y} \quad \text{is holomorphic on $\RR\times ]0,\beta[$}, \\
F_{x,y} \quad \text{is bounded continuous on $\RR\times[0,\beta],$}\\
F_{x,y}(t)=\omega(\sigma^\omega_{t}(y)x), \qquad t\in \RR, \\
F_{x,y}(i\beta+t)=\omega(x\sigma^\omega_{t}(y)), \qquad t\in \RR.
\end{gather*}

To a von Neumann algebra $\M$ on the Hilbert space $\H$, with a cyclic separating vector $\xi\in \H$, there is an associated
\emph{natural positive cone} $\P_\xi\subset \H$ def\/ined by $\P_\xi:=\{x\gamma_\omega(x)\xi \ | \ x\in \M\}^-$.
The natural cone is convex, self-dual, modularly stable $\Delta_\xi^{it}\P_\xi=\P_\xi$ for all $t\in \RR$, and pointwise stable under modular conjugation $J_\xi\zeta=\zeta$, for all $\zeta\in \P_\xi$.
The data $(\M,\H,J_\xi,\P_\xi)$ is usually called a \emph{standard form} for the von Neumann algebra $\M$.

{\sloppy
Let $(\M_j,\H_j,J_j,\P_j)$ for $j=1,2$ be two standard forms for the von Neumann algeb\-ras~$\M_1$,~$\M_2$.
Every $*$-isomorphism $\alpha:\M_1\to\M_2$ between two von Neumann algebras admits a unique unitary spatial implementation
$U_\alpha: \H_1\to\H_2$ such that $\alpha(x)=U_\alpha x U_\alpha^{-1}$, for all $x\in \M_1$,
$J_2=U_\alpha J_1U_\alpha^{-1}$ and $\P_2=U_\alpha (\P_1)$. In particular two standard forms of a von Neumann algebra are always naturally isomorphic and furthermore, the group of $*$-isomorphisms of a von Neumann algebra admits a unique unitary representation called \emph{standard implementation} with the three properties above.

}

\subsection{A categorical view of modular theory}\label{sec: m3}

It has long been known, mainly from the pioneering work of J.~Roberts~\cite{GLR}, that some of the main results in Tomita--Takesaki modular theory have a deep and enlightening interpretation in terms of $W^*$-categories.
Let us f\/irst introduce some def\/initions:

\begin{Definition}
A \emph{$*$-category} (also called \emph{dagger category} or \emph{involutive category})\footnote{See for example~\cite{BCL7} for more details and references.} is a category $\Cs$ with a contravariant functor $*:\Cs\to\Cs$ acting identically on the objects that is involutive i.e.~$(x^*)^*=x$ for all $x\in\Hom_ \Cs$.
A $*$-category is \emph{positive} if for all $x\in \Hom_\Cs(B,A)$ there is an element $z\in \Hom_\Cs(A,A)$ such that
$z^*z=x^*x$.
A $*$-category is an \emph{involutive inverse category} if $xx^*x=x$ for all $x\in \Hom_\Cs$. Whenever $\Hom_\Cs$ is equipped with a topology, we require the compositions and involutions to be continuous.
\end{Definition}
For a ($*$-)category $\Cs$ we will denote by $\Cs^o$ the set of identities $\iota_A$, for $A\in \Ob_\Cs$, simply by $\Cs$ the set
$\Hom_\Cs$ and by $\Cs^n$ the set of $n$-tuples $(x_1,\dots,x_n)$ of composable arrows $x_1,\dots,x_n\in \Hom_\Cs$.

\begin{Definition}
A \emph{$C^*$-category}\footnote{First introduced in P.~Ghez, R.~Lima, J.~Roberts~\cite{GLR} and further developed in
P.~Mitchener~\cite{Mit}.} is a positive $*$-category $\Cs$ such that: for all $A,B\in \Ob_\Cs$, the sets $\Cs_{AB}:=\Hom_\Cs(B,A)$ are complex Banach spaces; the compositions are bilinear maps such that $\|xy\|\leq\|x\|\cdot \|y\|$, for all $x\in \Cs_{AB}$, $y\in \Cs_{BC}$; the involutions are conjugate-linear maps such that $\|x^*x\|=\|x\|^2$, $\forall\, x\in \Cs_{BA}$.
A $C^*$-category is \emph{commutative} if for all $A\in \Ob_\Cs$, the $C^*$-algebras~$\Cs_{AA}$ are commutative.
\end{Definition}

There is a horizontally categorif\/ied version of Gel'fand--Na\u\i mark--Segal representation theo\-rem that works for states over
$C^*$-categories (see~\cite[Proposition~1.9]{GLR} and also~\cite[Theorem~2.8]{BRu}).
\begin{Definition}
A \emph{state} $\omega:\Cs\to\CC$ \emph{on a $C^*$-category} $\Cs$ is map $\omega:\Cs\to\CC$ that is linear when restricted to
$\Cs_{AB}$, for all $A,B\in \Ob_\Cs$, Hermitian i.e.~$\omega(x^*)=\cj{\omega(x)}$, for all $x\in \Cs$, normalized
i.e.~$\omega(\iota_A)=1_\CC$, for all $A\in \Ob_\Cs$, and positive i.e.~$\omega(x^*x)\geq0$, for all $x\in \Cs$.
\end{Definition}

\begin{Theorem}
Let $\omega$ be a state over the $C^*$-category $\Cs$. There is a representation $\pi_\omega$ of $\Cs$ on a~family $\H_A$ of Hilbert spaces indexed by the objects of $\Cs$ and there exists a family of normalized vectors $\xi_A\in \H_A$ such that
$\omega(x)=\ip{\xi_A}{\pi_\omega(x)\xi_B}_{\H_A}$, for all $x\in \Cs_{AB}$.
\end{Theorem}

Inspired by S.~Sakai def\/inition of $W^*$-algebra, P.~Ghez, R.~Lima, J.~Roberts~\cite[Def\/ini\-tion~2.1]{GLR} gave the following:
\begin{Definition}
A \emph{$W^*$-category} is a $C^*$-category $\Ws$ such that each space $\Ws_{AB}$ is the dual of a~Banach space $\Ws_{AB*}$ that is called the pre-dual of $\Ws_{AB}$.
\end{Definition}
A $W^*$-category can be equivalently def\/ined as a weakly closed $*$-subcategory of a $C^*$-category of $\Bf(\Hf)$ of bounded linear maps between Hilbert spaces in a given family $\Hf$.

Tomita--Takesaki modular theory admits a straighforward extension to this $W^*$-categorical setting, via semi-f\/inite faithful normal weights, resulting in a one-parameter group of object-preserving invertible $*$-functors, that has already been clearly investigated
in~\cite[Section~5]{GLR}. Some pleasant surprises, i.e.~Araki relative modular theory, appear as soon as one considers the spatial implementation of the modular $*$-functors as a horizontal categorif\/ication of the usual Tomita theory.

\begin{Theorem}
Given a normal faithful state $\omega$ over a $W^*$-category $\Ws$, there is an associated weakly continuous one-parameter group
$t\mapsto \sigma^\omega_t$ of $*$-autofunctors of $\Ws$ that in the Gel'fand--Na\u\i mark--Segal representation of $\Ws$ is spatially implemented by the one-parameter groups of unitaries $t\mapsto \Delta^{it}_{\omega_A,\omega_B}:\H_{\omega_B}\to\H_{\omega_B}$ i.e.~for all $x\in \Ws_{AB}$,
\begin{equation*}
\pi^\omega_{AB}(\sigma^\omega_t(x))=\Delta_{\omega_B,\omega_A}^{it}\pi^\omega_{AB}(x)\Delta^{-it}_{\omega_A,\omega_B}.
\end{equation*}
There is a family of conjugate-linear isometries $J_{\omega_A,\omega_B}: \H_{\omega_B}\to\H_{\omega_A}$ that spatially implements a conjugate-linear invertible contravariant isometric $*$-functor $\gamma_\omega:\pi^\omega(\Ws)\to\pi^\omega(\Ws)'$ i.e.
\begin{equation*}
\gamma_\omega(\pi^\omega_{AB}(x))=J_{\omega_B,\omega_A} \pi^\omega_{AB}(x)J_{\omega_B,\omega_A},
\end{equation*}
where $\pi^\omega(\Ws)$ is the von Neumann category represented on $\H_\omega$ and $\pi^\omega(\Ws)'$ denotes the commutant von Neumann category\footnote{Note that our def\/inition of commutant of a representation of a $W^*$-category does not coincide with the def\/inition given in~\cite[Section~4]{GLR}: there the commutant of a von Neumann category is a von Neumann algebra here it is a $W^*$-category. The Ghez--Lima--Roberts commutant is actually the $W^*$-envelope of our commutant.} of
$\pi^\omega(\Ws)$ that is defined by
\begin{equation*}
\pi^\omega_{AB}(\Ws)':=\{T\in \B(\H_\omega)_{AB} \ | \ T^*T\in \pi^\omega_{BB}(\Ws)', \ TT^*\in \pi^\omega_{AA}(\Ws)'\}.
\end{equation*}
\end{Theorem}
The positive operators $\Delta_{\omega_A,\omega_B}:\H_{\omega_B}\to\H_{\omega_B}$ are the \emph{Araki relative modular operators} associated to the pair of states $\omega_A,\omega_B$ and the operators
$J_{\omega_A,\omega_B}:\H_{\omega_B}\to\H_{\omega_A}$ are the \emph{Araki relative modular conjugations} for the pair
$\omega_A,\omega_B$.

\subsection{Weights, conditional expectations and operator valued weights}

Although, for simplicity, in this review we will mainly use (normal faithful) states on $\M$, it is also important (especially for the formulation of deeper results) to consider generalizations.

A \emph{weight} $\omega$ on a $C^*$-algebra $\A$ is a map $\omega:\As_+\to[0,+\infty]$ such that
$\omega(x+y)=\omega(x)+\omega(y)$ and $\omega(\alpha x)=\alpha\omega(x)$, for all $x,y\in \As_+$ and $\alpha\in \RR_+$.
A \emph{trace} is a weight that, for all $x\in\As$, satisf\/ies $\omega(x^*x)=\omega(xx^*)$.

The usual \emph{GNS-representation} associated to states admits a similar formulation in the case of weights.
To every weight $\omega$ on the $C^*$-algebra $\As$ there is a triple $(\H_\omega,\pi_\omega,\eta_\omega)$, where $\H_\omega$ is a Hilbert space, $\pi_\omega$ is a $*$-representation of $\As$ in $\B(\H_\omega)$ and $\eta_\omega:\Lg_\omega\to\H_\omega$ is a linear map with dense image def\/ined on the left ideal $\Lg_\omega:=\{x\in \As \ | \ \omega(x^*x)<+\infty\}$,
such that $\pi_\omega(x)\eta_\omega(z)=\eta_\omega(xz)$ and
$\omega(y^*xz)=\ip{\eta_\omega(y)}{\pi_\omega(x)\eta_\omega(z)}_{\H_\omega}$
for all $x\in \As$ and all $y,z\in \Lg_\omega$. A~GNS-representation for weights on a $C^*$-category is available as well.

A weight is \emph{faithful} if $\omega(x)=0$ implies $x=0$. A weight on a von Neumann algebra $\M$ is \emph{normal} if for every increasing bounded net in $\M_+$ with $x_\lambda\to x\in \M_+$ we have $\omega(x_\lambda)\to \omega(x)$ and it is
\emph{semi-finite} if the linear span of the cone $\Mg_{\omega+}:=\{x\in \M_+ \ \ \omega(x)<+\infty\}$ is dense in the
$\sigma$-weak operator topology in $\M$.

Tomita--Takesaki modular theory (i.e.~all the results mentioned in Sections~\ref{sec: m2}
and~\ref{sec: m3}) can be extended to the case of normal semi-f\/inite faithful weights on a von Neumann algebra and the formulation is essentially identical to the one already described in case of states. Natural positive cones are def\/ined also for weights $\omega$ as
$\P_\omega:=\{\pi_\omega(x)J_\omega\eta_\omega(x) \ | \ x\in \M\}^-$ and they enjoy the same properties already described in the previous Section~\ref{sec: m2}.

A von Neumann algebra $\M$ is \emph{semi-finite} if and only if it admits a normal semi-f\/inite faithful trace $\tau$.
In this case for every normal semi-f\/inite faithful weight $\omega$, the modular automorphism group $t\mapsto \sigma^\omega_t$ is inner i.e.~there exists a positive invertible operator $h$ af\/f\/iliated\footnote{This means that all the spectral projections of $h$ are contained in $\M$.} to $\M$ such that $\sigma^\omega_t(x)=h^{it}xh^{-it}$
for all $t\in \RR$ and $x\in \M$.

The following result is the celebrated Connes--Radon--Nikodym theorem.
\begin{Theorem}
Let $\phi$ be a normal semi-finite faithful weight on the von Neumann algebra $\M$.

For every other normal semi-finite faithful weight $\psi$ on $\M$, there exists a strongly continuous family $t\mapsto u_t$ of unitaries in $\M$ such that for all $x\in \M$ and all $t,s\in \RR$:
\begin{equation*}
\sigma^\psi_t(x)=u_t\sigma^\phi_t(x)u_t^*, \qquad u_{t+s}=u_t\sigma^\phi_t(u_s).
\end{equation*}
Furthermore, defining $\sigma^{\psi,\phi}_t(x):=u_t\sigma^\phi_t(x)=\sigma^\psi_t(x)u_t$, there exists a unique such family, denoted by $t\mapsto(D\phi : D\phi)_t$ for all $t\in \RR$,
and called the \emph{Connes--Radon--Nikodym derivative} of $\psi$ with respect to $\phi$,
that satisfies the following variant of the KMS-condition: there exists a bounded continuous function on
$\RR\times [0,1]$ analytic on $\RR\times]0,1[$ such that for all $x,y\in \Lg_\phi\cap\Lg_\psi^*$ and for all $t\in \RR$,
$f(t)=\psi(\sigma^{\psi,\phi}_t(x)y)$ and $f(t+i)=\phi(y\sigma^{\psi,\phi}_t(x))$.

If $t\mapsto u_t$ is a strongly continuous family of unitaries in $\M$ such that $u_{t+s}=u_t\sigma^\phi_t(u_s)$, for all $t,s\in \RR$, there exists a unique normal semi-finite faithful weight $\psi$ on $\M$ such that $(D\psi : D\phi)_t=u_t$, for all $t\in \RR$.
\end{Theorem}

The Connes--Radon--Nikodym derivatives satisfy the following properties for all normal semi-f\/inite faithful weights
$\omega_1$, $\omega_2$, $\omega_3$ on $\M$ and for all $t\in \RR$:
\begin{equation*}
(D\omega_1 : D\omega_2)_t \cdot (D\omega_2 : D\omega_3)_t = (D\omega_1 : D\omega_3)_t,
\qquad
(D\omega_1 : D\omega_2)_t^*=(D\omega_2 : D\omega_1)_t.
\end{equation*}

We also have the following fundamental theorem of A.~Connes.
\begin{Theorem}\label{th: sp-der}
Let $\M$ be a von Neumann algebra on the Hilbert space $\H$ and $\M'$ its commutant.
For any normal semi-finite faithful weight $\omega$ on $\M$ and any normal semi-finite faithful weight $\omega'$ on $\M'$,
there exists a positive operator $\Delta(\omega | \omega')$, the \emph{Connes' spatial derivative} of $\omega$ with respect to
$\omega'$ such that: $\Delta(\omega' | \omega)=\Delta(\omega | \omega')^{-1}$ and for all $t\in \RR$, 
\begin{gather*}
\sigma^\omega_t(x)=\Delta(\omega | \omega')^{it}x\Delta^{-it}(\omega | \omega') \quad \forall\, x\in \M,
\qquad
\sigma^{\omega'}_{-t}(y)=\Delta(\omega | \omega')^{it}y\Delta(\omega | \omega')^{-it} \quad \forall\, y\in \M'.
\end{gather*}
Furthermore, if $\omega_1$ and $\omega_2$ are normal semi-finite faithful weights on $\M$ we also have
\begin{equation*}
\Delta(\omega_2 | \omega')^{it}=(D\omega_2 : D\omega_1)_t\Delta(\omega_1 | \omega')^{it}, \qquad  \forall\, t\in \RR.
\end{equation*}
\end{Theorem}

A \emph{conditional expectation} $\Phi:\As\to\Bs$ from a unital $C^*$-algebra $\As$ onto a unital $C^*$-subal\-geb\-ra~$\Bs$ is a completely positive map\footnote{This means that for all $n\in \NN$, $\Phi^{(n)}:\MM_n(\As)\to\MM_n(\Bs)$ is positive, where $\MM_n(\As)$ denotes the unital $C^*$-algebra of $n\times n$ $\As$-valued matrices and $\Phi^{(n)}$ is obtained applying $\Phi$ to every entry.} such that $\Phi(b)=b$, $\Phi(x+y)=\Phi(x)+\Phi(y)$, $\Phi(b_1xb_2)=b_1\Phi(x)b_2$, for all $b,b_1,b_2\in \Bs$ and all $x,y\in \As$. By a theorem of J.~Tomiyama, $\Phi$ is a conditional expectation if and only if $\Phi$ is a~projection of norm one onto a subalgebra.
A conditional expectation is a~generalization of the notion of state that appears as long as we allow values to be taken in an arbitrary $C^*$-algebra in place of the usual complex numbers $\CC$.

Conditional expectations and modular theory are related by this result by M.~Takesaki.
\begin{Theorem}\label{th: cex}
Let $\N$ be a von Neumann subalgebra of the von Neumann algebra $\M$ and let $\omega$ be a normal semi-finite faithful weight on the von Neumann algebra $\M$ such that $\omega|_\N$ is semi-finite.
The von Neumann algebra $\N$ is modularly stable i.e.~$\sigma^\omega_t(\N)=\N$ for all $t\in \RR$, if and only if there exists a conditional expectation $\Phi:\M\to\N$ onto $\N$ such that $\omega\circ\Phi=\omega$.

Such conditional expectation is unique and normal.
\end{Theorem}

In the same way as weights are an ``unbounded'' version of states, we also have an ``unbounded'' version of conditional expectations. Here the role of real $\RR$ or positive real num\-bers~$\RR_+$ as possible values of a state, respectively weight, is taken by a von Neumann algebra $\N$ and its positive part $\N_+$ and the set $\widehat{\RR}_+:=[0,+\infty]$ of extended positive reals is replaced by $\widehat{\N}_+$, the \emph{extended positive cone} of $\N$, def\/ined as the set of lower semi-continuous maps $m:\M_{*+}\to[0,+\infty]$ such that $m(\phi+\psi)=m(\phi)+m(\psi)$, $m(\alpha\phi)=\alpha m(\phi)$,
$\forall\, \phi,\psi\in \M_{*+}$, $\alpha\in \RR_+$.

An \emph{operator valued weight} from the von Neumann algebra $\M$ to the von Neumann algebra~$\N$ is a map
$\Phi:\M_+\to \widehat{\N}_+$ taking values in the extended positive cone of $\N$ such that:
\begin{gather*}
\Phi(x+y)=\Phi(x)+\Phi(y), \qquad \Phi(\alpha x)=\alpha\Phi(x) \qquad \mbox{and}\\
 \Phi(u^*xu)=\Phi(x), \qquad \forall \, x,y\in \M_+, \quad \alpha\in \RR_+, \quad u\in \N.
\end{gather*}

With these def\/initions, Takesaki's theorem~\ref{th: cex} can be generalized as follows.
\begin{Theorem}\label{th: ovw}
The existence of a normal semi-finite faithful operator valued weight $\Phi$ onto a~subalgebra $\N$ of the von Neumann algebra $\M$ is equivalent to the existence of a pair of normal semi-finite faithful weights $\omega$ on $\N$ and
$\widetilde{\omega}$ on $\M$ such that
$\sigma^\omega_t(x)=\sigma^{\widetilde{\omega}}_t(x)$ for all $x\in \N$. There is a unique such $\Phi$ with the property $\widetilde{\omega}=\Phi\circ\omega$.
\end{Theorem}

\subsection{Connes--Takesaki duality}

The following version of the classical Connes--Takesaki duality theorem is making use of the Falcone--Takesaki canonical construction of the non-commutative f\/low of weights of a von Neumann algebra.

\begin{Theorem}\label{th: ct}
Let $\M$ be a von Neumann algebra. There exist a canonical one-parameter $W^*$-dynamical system, the \emph{non-commutative f\/low of weights}, $(\widetilde{\M},\theta)$ and a canonical normal $*$-morphism $\iota:\M\to\widetilde{\M}$ such that:
\begin{itemize}\itemsep=0pt
\item
the image of the canonical isomorphism coincides with the fixed points algebra of the dynamical system
i.e.~$\iota(\M)=\widetilde{\M}^\theta$,
\item
for every faithful semi-finite normal weight $\phi$ on $\M$ there is a canonical isomorphism of the $W^*$-dynamical system $(\widetilde{\M},\theta)$ with the $W^*$-dynamical system $(W^*(\M,\sigma^\phi),\widehat{\sigma}^\phi)$
induced by the dual action of $\sigma^\phi$ on the $W^*$-covariance algebra of $(\M,\sigma^\phi)$,
\item
there is a canonical operator valued weight $\Theta$ from $\widetilde{\M}$ onto $\iota(\M)$, given for all $x\in \widetilde{\M}_+$ by
$\Theta(x)=\int\theta_t(x)\,\text{d}t$, such that, for every faithful semi-finite normal weight $\phi$ on $\M$, the dual faithful semi-finite normal weight $\widetilde{\phi}:=\phi\circ \Theta$ on $\widetilde{\M}$ induces an inner modular automorphism group
i.e.~$\sigma^{\widetilde{\phi}}_{t}=\Ad_{e^{ik_\phi t}}$ with generator $k_\phi$ affiliated to $\widetilde{\M}$,
\item
there is a canonical faithful semi-finite normal trace $\tau$ on $\widetilde{\M}$ that is rescaling the one-parameter group $\theta$ i.e.~$\tau\circ\theta_t=e^{-t}\tau$, for all $t\in \RR$ and for all faithful semi-finite normal weights $\phi$ on $\M$ we have that
$\tau(x)=\widetilde{\phi}(e^{-k_\phi/2}xe^{-k_\phi/2})$, for all $x\in \widetilde{\M}_+$,
\item
for all faithful semi-finite normal weights $\phi$ on $\M$, we have that the $W^*$-dynamical system $(W^*(\widetilde{\M},\theta),\widehat{\theta})$ induced by the dual action of $\theta$ on the $W^*$-covariance algebra $W^*(\widetilde{\M},\theta)$ of $(\widetilde{\M},\theta)$ is canonically isomorphic with the $W^*$-dynamical system $(\M\otimes\B(L^2(\RR)),\sigma^\phi\otimes\rho)$, where
$\rho_t:=\Ad_{\lambda_{-t}}$ with $(\lambda_{t}\xi)(s):=\xi(s-t)$ the usual left regular action of $\RR$ on $L^2(\RR)$.
\end{itemize}
\end{Theorem}

\section{Modular theory in physics} \label{sec: mtp}

The history of the interplay between modular theory and physics is a very interesting and wide subject in itself that we are going to touch here only in a very simplif\/ied and incomplete way, mainly to provide a suitable motivation and background for the discussion of our proposals in the subsequent Section~\ref{sec: maqg}. For a more complete treatment, we refer the reader to the notes and remarks in O.~Bratteli,~D.~Robinson~\cite[Section~5]{BR}, the books by R.~Haag~\cite[Chapter~V]{H} and
G.~Emch~\cite[Chapter~10]{E} and to the recent expositions by  S.J.~Summers~\cite{Su1} and D.~Guido~\cite{G}.

The KMS condition for the characterization of equilibrium states, f\/irst introduced in quantum statistical mechanics by
R.~Kubo~\cite{Ku} and P.~Martin, J.~Schwinger~\cite{MS}, was reformulated in the algebraic quantum mechanical setting by
R.~Haag, N.~Hugenoltz, M.~Winnink~\cite{HHW} and its relation with Tomita modular theory was fully developed by M.~Takesaki~\cite{T1}.

Every physical state that satisf\/ies the $(\alpha,\beta)$-KMS condition on the $C^*$-algebra of observables is identif\/ied with an equilibrium state at temperature $1/\beta$ for the dynamics provided by the one-parameter group $\alpha$ of time-evolution of the system.
Every such state determines a unique dynamics of the observable algebra via its modular automorphism group.

The f\/irst indirect indications of the existence of a deep connection between (equilibrium) statistical mechanics (and hence modular theory), quantum f\/ield theory and gravity (that, after A.~Einstein's theory of general relativity, essentially means geometry of four-dimensional Lorentzian manifolds) came, after J.~Bardeen, B.~Carter, S.~Hawking~\cite{BCH} results on black hole laws, from the discovery of entropy of black holes by J.~Bekenstein~\cite{Be1,Be2}, black holes' thermal radiation by
S.~Hawking~\cite{Ha1,Ha2} and the vacuum thermalization ef\/fect by W.~Unruh~\cite{U}.
In practice, the vacuum state of a quantum theory of f\/ields on certain ``singular'' solutions of Einstein equation with horizons (black holes) presents a natural thermal behaviour manifested by entropy (proportional to the area of the horizon), thermal radiation and a temperature that is proportional to the acceleration of free falling observers at the horizon.

\looseness=1
G.~Gallavotti, M.~Pulvirenti~\cite{GP} have been discussing the role of Tomita--Takesaki theory for classical statistical mechanical systems. The point here is the existence of a correspondence between modular theory and von Neumann algebras on one side and Poisson geometry of classical systems on the other. This point of view has been further advocated by A.~Weinstein~\cite{W}.

The existence of an interplay between general relativity, gravitation and thermodynamics, has been reinforced by the important work of T.~Jacobson~\cite{Jac} that obtained for the f\/irst time a~thermodynamical derivation of Einstein equations from the equivalence principle. This work has been further expanded, among several authors, by T.~Padmanaban~\cite{Pa}.
This line of thoughts, has recently been exploited in order to infer that, being of thermodynamical origin, gravitation (contrary to electromagnetism and other subnuclear forces) cannot possibly be a~fundamental force of nature and hence should not be subjected to quantization, but explained as a macroscopic phenomenon emergent from a dif\/ferent theory of fundamental degrees of freedom (usually strings) and after the recent appearence of E.~Verlinde e-print~\cite{Ve} on the interpretation of Newtonian gravity as an entropic force (see the survey by S.~Hossenfelder~\cite{Ho} for a clean presentation of some of the directions and some related works) has led to a fantastic proliferation of research papers that is probably much more pertinent to address in a study of sociopathology of current scientif\/ic research.

More direct links between modular theory, quantum f\/ield theory and Minkowski space geometry appeared in the works of
J.~Bisognano, E.~Wichmann~\cite{BW1,BW2} (see for example D.~Guido~\cite[Section~3]{G} for a detailed discussion).
The vacuum state $\Omega$ of a scalar quantum f\/ield satisfying suitable G\aa rding--Wightman axioms and irreducibility, by Reeh--Schlieder theorem, is cyclic and separating when restricted to any von Neumann algebra of local observables $\Rs(\O)$ on a non-empty open region $\O\subset \MM^4$ of Minkowski space-time $\MM^4$ (with metric $\eta$ that we assume to be of signature
$(-1,+1,+1,+1)$) whose causal complement $\O':=\{x\in\MM^4 \ | \ \eta(x,z)>0, \ \forall\, z\in \O\}$ is non-trivial.
It follows that there is a natural one-parameter group of modular automorphisms induced by the vacuum state $\Omega$ on any local von Neumann algebra $\Rs(\O)$. Although the nature of such one-parameter groups, for generally shaped regions $\O$, is still now ``kind of'' mysterious, for the special case of space-time wedges $\W_{e_0,e_1}:=\{x\in\MM^4 \ | \ |\eta(x,e_0)|<\eta(x,e_1)\}$ (for any orthogonal pair of time-like $e_0$ and space-like $e_1$), the modular automorphism group $\Delta^{it}_\Omega$ of $\Rs(\W_{e_0,e_1})$ has a clear geometric implementation, as $\Delta^{-it}_\Omega=U(\Lambda_{\W_{e_0,e_1}}(t))$, via the one-parameter unitary group representing the unique one-parameter group of Lorentz boosts
$t\mapsto\Lambda_{\W_{e_0,e_1}}(t)$, in the plane generated by $e_0$, $e_1$, leaving invariant the wedge $\W_{e_0,e_1}$. Similarly the modular conjugation $J_\Omega$ is geometrically obtained as $J_\Omega=\Theta\circ U(R_{e_0,e_1}(\pi))$ via the unitary implemen\-ting the rotation $R_{e_0,e_1}(\pi)$ by $\pi$ in the space-like plane orthogonal to $e_0$, $e_1$ composed with the
conjugate-linear ``PCT operator'' $\Theta$.

In the case of a conformally covariant scalar f\/ield, P.~Hislop, R.~Longo~\cite{HL} have extended J.~Bisognano, E.~Wichmann results providing, via conformal transformations, a geometric implementation of the modular operators associated by the vacuum to the local von Neumann algebras of more general space-time regions, such as double cones and lightcones.

It is known that for massive free theories the vacuum does not act geometrically on the algebras of double cones (see for example
T.~Saf\/fary~\cite{Saf}) and that thermal states do not act geometrically even for wedge regions
(see H.-J.~Borchers, J.~Yngvason~\cite{BY}).

Further recent advances in the study of geometrical modular action induced for free Fermionic f\/ields by the vacuum on pairs of disjoint regions have been obtained by H.~Casini, M.~Huerta~\cite{CH} and by R.~Longo, P.~Martinetti, K.-H.~Rehren~\cite{LMR}. In this case the action is not completely geometrical and induce a form of ``dynamical mixing'' of the two regions.

The above mentioned results by W.~Unruh and J.~Bisognano, E.~Wichmann provided the basis for the work of G.~Sewell~\cite{Sew1,Sew2} that obtained a clear interpretation of S.~Hawking radiation.
For an observer in constant acceleration, whose world-line is described by the orbit of the boosts $\Lambda_{\W_{e_0,e_1}}(t)$, the border of the Minkoski space-time wedge $\W_{e_0,e_1}$ is a natural horizon (the observer will never be able to receive signals from events outside the wedge). The equivalence principle forces us to conclude that such an observer is (locally) equivalent to an observer in free-fall in a~gravitational f\/ield near the horizon and Reeh--Schlieder theorem provides us with an explanation of the Unruh thermalization of the vacuum state for such an observer. The vacuum state~$\Omega$, when restricted to the algebra of observables of the free-falling observer, becomes a~KMS state (equilibrium state), with respect to the natural time evolution, at a temperature that is proportional to the accelaration (i.e.~to the surface gravity at the horizon).

A formulation of J.~Bisognano, E.~Wichmann results in terms of Araki--Haag--Kastler axioms for algebraic quantum f\/ield theory was later achieved by H.-J.~Borchers~\cite{Bo1} opening the way to the development of several programs aiming at the reconstruction of (some aspects of) physical space-time geometry (mainly Poincar\'e symmetry) from the information encoded in the Tomita--Takesaki modular theory of some of the algebras of local observables and suitable conditions of ``modular covariance'' imposed on the net of observables.
\begin{itemize}\itemsep=0pt
\item[--]
In the theory of ``half-sided modular inclusions'' and ``modular intersections'' a (conformal or Poincar\'e) covariant net of von Neumann algebras satisfying J.~Bisognano, E.~Wichmann conditions is reconstructed from modular action conditions (encoding the positivity of the energy) imposed on pairs of von Neumann algebras with a common separating and cyclic vector. The main results in this direction have been obtained by
H.-J.~Borchers~\cite{Bo-mi}, \mbox{H.-W.}~Wiesbrock~\cite{Wi1,Wi2}, R.~K\"ahler, H.-W.~Wiesbrock~\cite{KW}, H.~Araki, L.~Zsido~\cite{AZ} (see the survey by H.-J.~Borchers~\cite{Bo2} for more details and references).
\item[--]
In the works of R.~Brunetti, D.~Guido, R.~Longo~\cite{BGL-coh,GL} ``modular covariance'', i.e.~the J.~Bisognano, E.~Wichmann condition of geometric implementation for the one-parameter modular groups associated to von Neumann algebras of wedge regions for a local net in Minkowski space-time, entails a covariant representation of the full Poincar\'e group.
\item[--]
In the ``geometric modular action'' program proposed by D.~Buchholz, S.J.~Summers~\cite{BuS1,BuS2} and developed in a series of  works by D.~Buchholz, O.~Dreyer, M.~Florig, S.J.~Summers, R.~White~\cite{BDFS,BuS3,BFS,Su2,SuW}, a condition of modular action for the modular conjugations associated to a given state on the von Neumann algebras of suitable regions of space-time, allows to reconstruct the full unitary dynamics of the theory, the isometry group of space-time and its covariant action on the net of observables. A further condition of ``modular stability'' imposed on the one-parameter modular groups assures the positivity of the energy.
In some cases (including Minkowski space-time) a remarkable complete reconstruction of space-time from the vacuum state and the algebras of local observables has been achieved~\cite{SuW,Su2}.
\item[--]
In the ``modular localization program'' by R.~Brunetti, D.~Guido, R.~Longo~\cite{BGL} (see also
B.~Schroer, H.-W.~Wiesbrock~\cite{Sc1,Sc2,SW1,SW2}, F.~Lled\'o~\cite{Lle1}, J.~Mund, B.~Schroer, J.~Yngvason~\cite{MSY},
J.~Mund~\cite{Mu2} and references therein), assuming the existence of a~representation of the Poincar\'e group on the one-particle Hilbert space, and exploiting the modular operators associated to wedge regions, one can reconstruct a covariant net of von Neumann algebras of free-f\/ields on Minkowski space-time. A modular reconstruction of space-time via ``modular positions'' has been conjectured by
N.~Pinamonti~\cite{Pi}.

An extremely interesting perspective on the historical and conceptual development of the notion of ``modular localization'' intrinsic in quantum f\/ield theory, emphasizing its radical dif\/ferences with the usual ``Born--Newton--Wigner localization'' in quantum mechanics can be found in the recent papers by B.~Schroer~\cite{Sc3,Sc4,Sc5,Sc6,Sc7,Sc8}.
\item[--]
In the ``modular nuclearity program'' (see R.~Haag~\cite[Section~V.5]{H} and references therein) since the work of D.~Buchholz, C.~D'Antoni, R.~Longo~\cite{BDL} the modular theo\-ry induced by the vacuum state $\Omega$ on a local von Neumann algebra $\Rs(\O)$ of a bounded open non-empty region $\O$ is related with geometrical properties of phase-space (locally f\/inite degrees of freedom) via the compactness/nuclearity of the map $x\mapsto \Delta_\Omega^\lambda x\Omega$, for all $x$ in a local subalgebra of $\Rs(\O)$ and for $0<\lambda<1/2$.
\item[--]
In the ``form factor program'' initiated by B.~Schroer~\cite{Sc2} modular theory plays a basic role in the reconstruction of local f\/ields from the scattering matrix. With the input of modular nuclearity D.~Buchholz, G.~Lechner~\cite{BL,Le1,Le2,Le3,Le4,Le5} obtained
non-trivial algebras of local observables and suggested the possibility to make use of algebraic quantum f\/ield theory for the construction of explicit models of interacting f\/ields (see D.~Buchholz, S.J.~Summers~\cite{BuS3}).
\end{itemize}

One should also notice that
a remarkable proof of J.~Bisognano, E.~Wichmann theorem in algebraic quantum f\/ield theory, based on scattering assumptions of asymptotic completeness and massive gap, has been f\/inally obtained by J.~Mund~\cite{Mu1} and further generalized to the case of massive particles with braid group statistic~\cite{Mu3,Mu4} in three dimensional space-time.

Modular theory has been explicitly invoked by A.~Connes, C.~Rovelli~\cite{CR} in order to give a~precise mathematical implementation of C.~Rovelli's ``thermal time hypothesis'' (see C.~Ro\-vel\-li~\cite[Sections~3.4 and~5.5.1]{Ro1} and also
C.~Rovelli and M.~Smerlak~\cite{RS2}) according to which the macroscopic f\/low of classical time is induced by a statistical state on the algebra of physical observables for a relativistically covariant system (that is intrinsically timeless).
In this case the specif\/ication of an ``equilibrium state'' as a KMS-functional on the $C^*$-algebra of observables, via Tomita--Takesaki modular theory, induces a unique one-parameter group that is interpreted as a state dependent macroscopic time-evolution.
At the level of von Neumann algebras the time-f\/low is unique modulo inner automorphisms i.e., in contrast to $C^*$-algebras, von Neumann algebras have an intrinsic macroscopic dynamics. Relations between the thermal time hypothesis and modular covariance have been elaborated by P.~Martinetti~\cite{Mart1,Mart3} and P.~Martinetti, C.~Rovelli~\cite{MR}.

Intriguing proposals to make use of Tomita--Takesaki modular theory to introduce a ``micro-macro duality'', explaining the appearence of macroscopic degrees of freedom from microscopic observables via a generalization of superselection theory, has been put forward by I.~Ojima~\cite{O1,O2} and I.~Ojima, M.~Takeori~\cite{OT}. This is, to our knowledge, the f\/irst tentative physical application of Connes--Takesaki duality in physics.

In their investigation of the role of exotic dif\/ferential structures in four dimensional space-time for a quantum gravity theory, T.~Asselmeyer-Maluga, J.~Krol~\cite{AK} have recently made use of type III von Neumann algebras.

\section{Non-commutative geometry} \label{sec: ncg}

Non-commutative geometry (in A.~Connes' sense) is in essence the most powerful modern incarnation of R.~Decartes's idea of ``algebraization of geometry'' i.e.~the codif\/ication of the geometry of spaces via algebras of their coordinates functions.
Its foundations are based on the existence of dualities (contravariant equivalences) between some specif\/ic categories of geometrical spaces and suitable categories of commutative algebras that, upon the removal of the commutativity axiom, provide a possible  def\/inition of (algebras of functions on) ``non-commutative spaces''\footnote{For an introduction to the exciting f\/ield of non-commutative geometry, the reader can already utilize several textbooks and lectures notes such as: G.~Landi~\cite{Lan},
H.~Figueroa, J.~Gracia-Bond\'{\i}a, J.~V\'arilly~\cite{FGV}, J.~V\'arilly~\cite{Var}, M.~Khalkhali~\cite{Kha,Kha2}, A.~Rennie~\cite{Re2}, apart from the original sources A.~Connes~\cite{C2} and A.~Connes, M.~Marcolli~\cite{CM2}. Several older, but still interesting introductory expositions to non-commutative geometry with a physicist's perspective are in R.~Coquereaux~\cite{Co1,Co2}, J.~Gracia-Bond\'{\i}a, J.~V\'arilly~\cite{GV}.}.

The most basic example of such dualities is the celebrated \emph{Gel'fand--Na\u\i mark duality} between the category of continuous functions of compact Hausdorf\/f topological spaces and the category of unital $*$-homomorphisms of commutative unital $C^*$-algebras,  opening the way for the consideration of unital non-commutative $C^*$-algebras as ``non-commutative compact Hausdorf\/f topological spaces''\footnote{See for example N.~Landsman~\cite[Section~6]{La} or, for a horizontally categorif\/ied version, our
papers~\cite{BCL5,BCL6}.}.

At the (topological) measure theoretic level, we have a duality between measurable maps of f\/inite Borel measure spaces and unital measure preserving $*$-homomorphisms of commutative unital $C^*$-algebras with a Radon measure that provide the basis for the consideration of states (or more generally weights) on unital $C^*$-algebras as ``non-commutative Radon measures''.

Along these directions, at the smooth metric level, the most successful result obtained so far is Connes' reconstruction theorem~\cite{C9,C10} that seems to indicate the existence of a duality between f\/inite-dimensional compact connected orientable Riemannian spin manifolds (with given irreducible spinorial bundle and spinorial charge conjugation) and a specif\/ic class of spectral triples with real structure.
The best candidate for a class of morphisms of spectral triples supporting this duality consists of certain smooth correspondences, recently proposed by B.~Mesland~\cite{Mes}.

\subsection{Spectral triples} \label{sec: st-ac}

Apart from technicalities, a spectral triple is a ``special kind of quantum dynamical system'' represented on a Hilbert space and, as already noted by several people (see M.~Paschke~\cite{P3} for example and, for a point of view closer to ``functorial quantum f\/ield theory'', also U.~Schreiber~\cite{Sch1,Sch2}), its def\/inition strongly and intriguingly resembles the axioms of algebraic quantum mechanics. Specif\/ically, a \emph{compact spectral triple} $(\As,\H, D)$ is given by:
\begin{itemize}\itemsep=0pt
\item
a unital pre-$C^*$-algebra $\As$ i.e.~an involutive normed unital algebra whose norm closure is a $C^*$-algebra,
\item
a faithful representation $\pi:\As\to\B(\H)$ of $\As$ in the von Neumann algebra $\B(\H)$ of bounded linear operators on a Hilbert space $\H$,
\item
a one-parameter group of of unitaries whose generator $D$, the Dirac operator, is such that
\begin{itemize}\itemsep=0pt
\item
the domain $\dom(D)$ is invariant under all the operators $\pi(a)$, with $a\in \As$,
\item
all the commutators $[D,\pi(a)]_-:=D\circ \pi(a)-\pi(a)\circ D$, def\/ined on $\dom(D)$, can be extended to bounded linear operators on
$\H$,
\item
the resolvent $(D-\mu I)^{-1}$ is compact for all $\mu\notin\Sp(D)$.
\end{itemize}
\end{itemize}
As revealed by these last conditions, one of the main motivations in the development of spectral triples comes from the index theory for elliptic operators and the realization of J.~Kasparov $K$-homology via unbounded Fredholm modules: by results of
S.~Baaj, P.~Julg~\cite{BJ} a spectral triple determines a pre-Fredholm module $(\As,\H,D(1+D^2)^{-1/2})$.
\begin{itemize}\itemsep=0pt
\item
A spectral triple is called \emph{even} if there exists a bounded self-adjoint operator $\Gamma \in \B(\H)$, called a \emph{grading operator}, such that:
\begin{gather*}
\Gamma^2=\text{Id}_\H; \qquad  [\Gamma, \pi(a)]_{-}=0, \quad \forall \, a \in \As; \qquad
[\Gamma, D]_{+}=0,
\end{gather*}
\item
a spectral triple is \emph{$\theta$-summable} if the operator $\exp(-t D^2)$ is a trace-class operator for all $t>0$, and it is
\emph{$n$-dimensional} if the Dixmier trace of $|D|^{-n}$ is f\/inite nonzero for a certain $n\in \NN$.
\end{itemize}

Some further conditions on spectral triples are motivated more directly from the spin geometry of Clif\/ford bundles and Atiyah--Singer Dirac operators and can be recasted in terms of the following \emph{non-commutative Clifford algebra}\footnote{A justif\/ication for such a name comes from the fact that, for commutative spectral triples, $\Omega_D(\As)$ reduces to the usual Clif\/ford algebra of the manifold represented on the Hilbert space of square integrable sections of the spinor bundle. Of course, in the non-commutative case such denomination might be misplaced and, at least in the real case, it would probably be better to use the larger algebra
$\Omega_{D,J}(\As):=\spa\{\Omega_D(\As)\cup J\Omega_D(\As)J\}$.}
$\Omega_D(\As)\subset \B(\H)$
\begin{equation*}
\Omega_D(\As):=\spa \{\pi(a_0)[D,\pi(a_1)]_- \cdots [D,\pi(a_n)]_- \  |
\ \  n\in \NN, \ a_0, \dots, a_n \in \As\},
\end{equation*}
where we assume that for $n=0 \in \NN$ the term in the formula simply reduces to $\pi(a_0)$.

\begin{itemize}
\item
a spectral triple is said to be \emph{regular} if, for all $x\in \Omega_D(\As)$, the functions given by
\[
\Xi_x: \ t\mapsto \exp(it|D|)x\exp(-it|D|)
\]
are regular, i.e.~$\Xi_x\in \text{C}^\infty(\RR, \B(\H))$.
\item
A spectral triple is \emph{finite} if $\H_\infty := \cap_{k=1}^\infty \text{Dom}\, D^k$ is a f\/inite projective $\As$-bimodule
and \emph{absolutely continuous} if, there exists an Hermitian form $(\xi,\eta)\mapsto\mip{\xi}{\eta}$ on $\H_\infty$ such that, for all $a\in \As$, $\ip{\xi}{\pi(a)\eta}$ is the Dixmier trace of $\pi(a)\mip{\xi}{\eta}|D|^{-n}$,
\item
an $n$-dimensional spectral triple is said to be \emph{orientable} if in the non-commutative Clif\/ford algebra
$\Omega_D(\As)$ there is a volume element
$\sum_{j=1}^m\pi(a^{(j)}_0)[D, \pi(a^{(j)}_1)]_-\cdots[D,\pi(a^{(j)}_n)]_-$  that coincides with the grading operator $\Gamma$ in the even case or the identity operator in the odd case,\footnote{In the following, in order to simplify the discussion, we will always refer to a ``grading operator'' $\Gamma$ that actually coincides with the grading operator in the even case and that is by def\/inition the identity operator in the odd case.}
\item
a spectral triple $(\As,\H, D)$ satisf\/ies \emph{Poincar\'e duality} if the $C^*$-module completion of
$\H_\infty$ is a Morita equivalence bimodule between (the norm completions of) $\As$ and $\Omega_D(\As)$ (see
A.~Rennie, J.~V\'arilly~\cite{RV1,RV2} for details),
\item
a spectral triple is said to have a \emph{real structure} if there exists an anti-unitary operator $J: \H \to \H$ such that:
\begin{gather*}
[\pi(a), J\pi(b^*)J^{-1}]_{-}=0, \qquad \forall\, a,b \in \As; \\
[\, [D, \pi(a)]_{-}, J\pi(b^*)J^{-1}]_{-}=0, \qquad \forall \, a,b \in \As, \qquad {\emph{first order condition};} \\
J^2=\pm\text{Id}_\H;  \qquad [J,D]_{\pm}=0;
\quad \text{and, only in the even case,} \qquad
[J,\Gamma]_{\pm}=0,
\end{gather*}
where the choice of $\pm$ in the last three formulas depends on the ``dimension'' $n$ of the spectral triple modulo $8$ in accordance to  the following table:
\begin{center}\label{tb: J}
\begin{tabular}{|l|c|c|c|c|c|c|c|c|}
\hline
$n$	&$0$	&$1$	&$2$	&$3$	&$4$	&$5$	&$6$	&$7$	\\
	\hline
$J^2=\pm\text{Id}_\H$	&	$+$	&	$+$		&	$-$		&	$-$		&	$-$		&	$-$		&	$+$		& $+$		 \\
	\hline
$[J,D]_{\pm}=0$				&	$-$	&	$+$		&	$-$		&	$-$		&	$-$		&	$+$		&	$-$		& $-$		 \\
\hline
$[J,\Gamma]_{\pm}=0$	&	$-$	&			&	$+$		&			&	$-$		&			&	$+$		& 		\\
\hline
\end{tabular}
\end{center}
\end{itemize}
As we already said, the paradigmatic examples of such a structure are the \emph{Atiyah--Singer spectral triples} coming from the theory of Atiyah--Singer Pauli--Dirac operators on spin manifolds. More precisely, given any $n$-dimensional compact orientable (connected) spinorial Riemannian manifold $M$, with a given spinor bundle $S(M)$ and a given spinorial charge conjugation
$C: S(M)\to S(M)$, there is an associated $n$-dimensional orientable f\/inite absolutely continuous regular real spectral triple that satisf\/ies Poincar\'e duality $(\As_M,\H_M,D_M)$ where
$\As_M:=C^\infty(M)$ is the pre-$C^*$-algebra of smooth complex-valued functions on $M$; $\H_M$ is the completion of the
$\As_M$-module of smooth sections of the spinor bundle under the inner product
$\ip{\sigma}{\tau}:=\int_M\ip{\sigma(p)}{\tau(p)}_p \,\text{d}\mu_M(p)$ induced by the volume form $\mu_M$; and $D_M$ is the Atiyah--Singer Pauli--Dirac operator def\/ined on $\H_M$ as the closure of the contraction of the spinorial connection induced by the Levi-Civita connection with the Clif\/ford multiplication (see the discussion in~\cite[Section~3.2.1]{BCL6} for more details).

Several examples of spectral triples are now available (a non-exhaustive list with relevant references can be found
in~\cite[Section~3.3]{BCL6}) although in some situations (such as for spectral triples associated to quantum groups, for non-compact spectral triples) a few of the preceeding axioms have to be modif\/ied or weakened.

Among the spectral triples that are ``purely quantal'', i.e.~are not directly related to (deformations of) classical spaces, we explicitly mention, for its relevance in later discussion, the construction of
\emph{Antonescu--Christensen spectral triples for AF $C^*$-algebras}~\cite{AC}.
\begin{Theorem}\label{th: ac}
Given a filtration of unital finite dimensional $C^*$-algebras
\begin{equation*}
\As_0:=\CC 1_\As\subset \As_1 \subset \cdots \subset \As_n\subset \As_{n+1}\subset \cdots
\end{equation*}
and a faithful state $\omega$ on the inductive limit of the filtration
$\As:=(\cup_{n=1}^{+\infty} \As_n)^-$ with GNS representation $(\pi_\omega,\H_\omega,\xi_\omega)$, denote by
$P_n\in \Bs(\H_\omega)$ the othogonal projection onto $\pi_\omega(\As_n)\xi_\omega$, by $E_n:=P_n-P_{n-1}$
$($we assume $E_0:=P_0)$ and by $\theta_n$ the continuous projection of $\As$ onto $\As_n$
$($that satisfies $\theta_n(a)\xi_\omega=P_n a\xi_\omega)$.
For any sequence $(\beta_n)$ such that $\sum_{n=1}^{+\infty}\beta_n<+\infty$ and any sequence $(\gamma_n)$ such that
$\|\theta_n(a)-\theta_{n-1}(a)\| \leq \gamma_n\|E_na\xi_\omega\|$ for all $a\in \As$, there is a family of spectral triples
$(\As,\H_\omega,D_{(\alpha_n)})$, indexed by a sequence of positive real numbers $(\alpha_n)$,
with $D_{(\alpha_n)}:=\sum_{n=1}^{+\infty}\alpha_n E_n$ and $\alpha_n:=\gamma_n/\beta_n$.
\end{Theorem}

The functorial properties of the Antonescu--Christensen construction (via suitable def\/initions of morphisms of f\/iltrations of AF
$C^*$-algebras and morphisms of spectral triples) is an interesting subject that will be addressed in a forthcoming work.\footnote{Bertozzini~P., Conti R., Lewkeeratiyutkul~W., Morphisms of spectral triples and AF algebras, in preparation.}

\subsection{Other spectral geometries}\label{sec: other}

There are non-commutative spectral geometries that do not f\/it exactly in the axiomatization of A.~Connes spectral triples.
Some of these geometries are just modif\/ications or extensions of the spectral triple framework (such as Lorentzian spectral triples); others describe situations in which second order operators, such as Laplacians, are in place of the f\/irst order Dirac operator.
In some cases the proposed axioms are making an even more direct appeal to Tomita--Takesaki modular theory (for example
S.~Lord's non-commutative Riemannian geometries and J.~Fr\"ohlich's quantized phase spaces) and so they might become particularly signif\/icant later on in order to interpret the meaning of modular spectral geometries that we will introduce in
Section~\ref{sec: maqg}.

\emph{Lorentzian spectral triples} have been originally proposed by A.~Strohmaier~\cite{Str} and their study has been carried on by W.~Van Suijlekhom~\cite{Sui}, M.~Paschke, A.~Sitarz~\cite{PS}, M.~Paschke, A.~Rennie, R.~Verch~\cite{PRV,PV2} and M.~Borris, R.~Verch~\cite{BV}.
The main motivation arises from the Dirac operator acting on the Kre\u\i n space of square integrable sections of a given spinor bundle (with charge conjugation) on a spinorial Lorentzian (or more generally semi-Riemannian) manifold.
Clearly all the dif\/ferential geometric motivations underlying the construction of spectral triples are still there, but the input from index theory is not so easily readable (for example there is a condition on the compactness of the resolvent of $(D^*D+DD^*)^{1/2}$, where $*$ denotes the Kre\u\i n space adjoint).
Since we are not making immediate usage of this notion here, we refer the reader to the more recent def\/inition available in M.~Paschke, A.~Sitarz~\cite[Section~2]{PS}.

The f\/irst axiomatization of \emph{non-commutative Riemannian manifolds}, that are not necessarily spinorial, has been put forward by J.~Fr\"ohlich, O.~Grandjean, A.~Recknagel~\cite[Section~2.2]{FGR4}.
They initially introduce the following set of f\/ive data $(\As,\H,d,\gamma,\star)$ that they call ``$N=(1,1)$ spectral data'' where:

\smallskip

$\As$ is an involutive unital algebra faithfully represented in $\B(\H)$,

\smallskip

$d$ is a closed densely def\/ined operator on $\H$ such that:
\begin{itemize}\itemsep=0pt
\item
$d \circ d = 0$,
\item
$[d,\pi(x)]_-$ extends to a bounded linear operator on $\H$ for all $x\in \As$,
\item
def\/ining $H:=d^*d+dd^*$, the operator $\exp(tH)$ is trace class for all $t\in \RR$,
\end{itemize}

$\gamma$ is a grading operator such that $\gamma=\gamma^*$, $\gamma\circ\gamma=I$, $[\gamma,\pi(x)]_-=0$, for all $x\in \As$ and $[\gamma,d]_+=0$,

\smallskip

$\star$ is a unitary operator on $\H$ such that $[\star,\pi(x)]_-=0$, for all $x\in \As$ and $\star\circ d=-1d^*\circ\star$,
(actually in place of the factor $-1$ in this last condition J.~Fr\"ohlich and collaborators admit a~more general phase $\zeta\in \TT$).

The basic motivating example here comes form the algebra of smooth functions acting on the Hilbert space of square integrable dif\/ferential forms on a Riemannian manifold, with $d$ being the exterior dif\/ferential and $\star$ the Hodge duality operator.

Then, to characterize \emph{non-commutative manifolds}, they further impose the existence of a~self-adjoint operator $T$ with integral spectrum, with $\gamma=f(T)$ where $f(\pm n)=(-1)^n$, such that $[T,\pi(x)]_-=0$, for all $x\in \As$, and $[T,d]=d$.
Def\/ining $D:=d+d^*$ and $\cj{D}:=i(d-d^*)$, they consider the ``non-commutative Clif\/ford algebra'' $\Omega_D(\As)$ and impose the existence of a cyclic separating vector for the weak closure of $\Omega_D(\As)''$ and the boundedness of all the commutators
$[\cj{D},J_\xi\pi(x)J_\xi]_-$, for $x\in \As$. Finally, denoting by $\Omega_{\cj{D}}(\cdot)^\pm$ the even/odd parts under the grading~$\gamma$, they require the condition
$J_\xi\Omega_D(\As)J_\xi=\Omega_{\cj{D}}(J_\xi\pi(\As)J_\xi)^+\oplus \gamma\Omega_{\cj{D}}(J_\xi\pi(\As)J_\xi)^-$.

A non-commutative manifold $(\As,\H,d,\star,T,\xi)$ is said to be \emph{non-commutative spin$^c$} if there is a factorization
$\H=\H_D\otimes_{Z(\M)} \H_{\cj{D}}$ of $\H$ over the center of $\M$ with a left module for $\Omega_D(\As)$ and
right module for $J_\xi\Omega_D(\As)J_\xi$.

Finally, in what is the f\/irst full use of Tomita--Takesaki modular theory in non-commutative geometry,
J.~Fr\"ohlich, O.~Grandjean, A.~Recknagel propose a def\/inition  \emph{non-commutative phase-space}.
On the algebra $\Omega_D(\As)$, obtained from a $N=(1,1)$ spectral data $(\As,\H,d,\gamma,\star)$, they consider a
$\beta$-KMS state given by
$\phi_\beta(x):=\text{Tr}_\H (x\exp(\beta D^2))/\text{Tr}_\H (\exp(\beta D^2))$, $\forall \beta>0$.

In the $\phi_\beta$-GNS-representation $(\H_\beta,\pi_\beta,\xi_\beta)$, they note that every bounded operator $L\in\B(\H)$ determines an operator $L_\beta$ on $\H_\beta$, via the sesquilinear form $\phi_\beta(x^*Ly)$ and use this to def\/ine the commutators
$-i[\cj{D},J_\beta\pi_\beta(x)J_\beta]_-$ for all $x\in \As$, via the derivative at $t=0$ of the map
$t\mapsto \exp(it\cj{D})J_\beta\pi_\beta(x)J_\beta\exp(-it\cj{D})$. Imposing the boundedness of
$[\cj{D},J_\beta\pi_\beta(x)J_\beta]_-$ they f\/inally require
$J_\beta\Omega_D(\As)J_\beta=\Omega_{\cj{D}}(J_\beta\pi(\As)J_\beta)^+\oplus \gamma_\beta\Omega_{\cj{D}}(J_\beta\pi(\As)J_\beta)^-$.

Another related proposal for axiomatizations of \emph{non-commutative Riemannian geometries} by S.~Lord~\cite{Lo} is closer in spirit to A.~Connes' def\/inition of spectral triples and invokes directly the standard form of von Neumann algebras: given a K-cycle
$(\As,\H,D)$ (with appropriate smoothness and summability conditions) it is assumed the existence of a cyclic separating vector
$\xi\in\H$ for the weak closure of the non-commutative Clif\/ford algebra $\Omega_D(\As)''\subset\B(\H)$ and conditions are further required in order to assure that $J_\xi$ provides a real structure and that absolute continuity holds for $(\As,\H,D)$. As a consequence, the representation of $\As$ on $\H$ coincides with the GNS-representation induced by the trace given by the non-commutative integral.
The basic commutative example is again the triple $(C^\infty(M),L^2(\Lambda_\bullet(M)),d+d^*)$ for a compact Riemannian (not necessarily spinorial) manifold $M$, where $\Lambda_\bullet(M)$ denotes the exterior bundle of $M$ and $d+d^*$ is the usual Dirac operator on forms. Unfortunately the reconstruction theorem for such Riemannian manifolds (in the commutative case) proved under the additional assumptions of f\/initeness, orientability, Poincar\'e duality, (being based on a previous incomplete proof of Connes' reconstruction theorem by A.~Rennie~\cite{Re1}) is not yet secured.

\section{Modular theory in non-commutative geometry} \label{sec: mncg}

The f\/irst link between A.~Connes' non-commutative dif\/ferential geometry and (semi-f\/inite) modular theory (as stated in~\cite{CPR4}) seems to be in a work by A.~Connes, J.~Cuntz~\cite{CC} on the relation between cyclic cocycles and semi-f\/inite Fredholm modules.

Anyway modular operators in relation to spectral triples appeared explicitly for the f\/irst time in a kind of mysterious way in
A.~Connes' papers~\cite{C3,C4} on the def\/inition of real structure for a spectral triple and was motivated by the presence of modular anti-isometries in specif\/ic f\/inite dimensional algebras arising in the study of the standard model in particle physics as well as from the charge conjugation operators induced on the spinorial module from the Tomita theory coming from the canonical trace on the Clif\/ford algebra of a spin manifold.

Modular theory appears again, but always only through the presence of modular conjugations, in the axiomatization of
non-commutative Riemannian manifolds and even more explicitly in the non-commutative phase-spaces proposed by
J.~Fr\"ohlich, O.~Grandjean, A.~Recknagel~\cite[Sections~5.2.6]{FGR3,FGR4}, that we recalled in the previous Section~\ref{sec: other}.

A similar prominent role was taken by Tomita--Takesaki modular theory in the tentative axiomatization of non-commutative Riemannian manifolds by S.~Lord~\cite{Lo} that also conjectured~\cite[Section VII.3]{Lo} the existence of a more elaborated ``non-commutative geometrical theory'' relating the modular f\/low for von Neumann algebras in standard form with A.~Connes' spectral triples. A similar point of view has been stated by M.~Paschke, R.~Verch~\cite[Section~6]{PV1}.

The steps that have been leading to the most successful current approach to semi-f\/inite non-commutative geometry and beyond are clearly described in the beautiful recent notes by A.~Carey, J.~Phillips, A.~Rennie~\cite{CPR4} and are essentially motivated by extensions of  classical index theory. Starting from the development of a Fredholm theory for von Neumann algebras by
M.~Breuer~\cite{Br1,Br2}, a f\/irst def\/inition of spectral triple with respect to a semi-f\/inite von Neumann algebra has been proposed by
M.-T.~Benameur, T.~Fack~\cite{BF} and the theory has been further developed~\cite{CPS1,CPS2,CP} in a monumental series of works by M.-T.~Benameur, A.~Carey, J.~Phillips, R.~Nest, A.~Rennie, A.~Sedaev, F.~Sukochev, K.~Tong, K.~Wojciechowski that described local index formulas for semi-f\/inite spectral triples~\cite{BCPRSW,CPRS1,CPRS2,CPRS3,CPRS4} (in this context see also the recent work by A.Lai~\cite{Lai}), an integration~\cite{CRSS,CGRS} and a
twisted version of $KK$-theory adapted to the situation at
hand~\cite{CPR2,CNNR,CRT,KNR} and developed crucial examples leading to a def\/inition of ``modular spectral triples'' for von Neumann algebras equipped with a periodic action of the modular group of a KMS
state~\cite{PaR, CPR1, CPR3, CPR4, CPPR}.

A.~Connes, H.~Moscovici~\cite{CMo,Mos} have recently put forward an alternative approach to the non-commutative geometry for type III von Neumann algebras where the notion of spectral triple is modif\/ied with a ``twist'' in order to retain the usual properties of pairing with $K$-theory (see also F.~D'Andrea~\cite{D} for further applications to spectral triples for quantum groups).

Notable examples of semi-f\/inite spectral triples have already appeared in physics via the work of
J.~Aastrup, J.~Grimstrup~\cite{AG1,AG2,AG3} linking A.~Connes' spectral triples and loop quantum gravity
(see the papers by A.~Aastrup, J.~Grimstrup, R.~Nest, M.~Paschke~\cite{AGN1,AGN2,AGN3,AGN4,AGNP,AGP,AGP2}).
It is also interesting to remark (see the end of Subsection~\ref{sec: ac} and the f\/inal Subsection~\ref{sec: otherqg}) that the
semi-f\/inite spectral triples def\/ined by J.~Aastrup, J.~Grimstrup, R.~Nest appear to be related to a generalization of the
C.~Antonescu, E.~Christensen construction of spectral triples (where the condition of f\/inite dimensionality of the eigenspaces of~$D$ is relaxed).

\subsection{Modular spectral triples (after Carey--Phillips--Rennie)} \label{sec: mst}

Apart from some irrelevant uniformization of notation, the following are the def\/initions adopted in
A.~Carey, J.~Phillips, A.~Rennie~\cite{CPR1,CPR4, CPPR}.

\begin{Definition}\label{def: s-st}
A \emph{semi-finite spectral triple} $(\As,\H,D)$ relative to a normal semi-f\/inite faithful trace $\tau$ on a semi-f\/inite von Neumann algebra $\Ns$, is given by:
\begin{itemize}\itemsep=0pt
\item
a faithful representation $\pi:\As\to \Ns\subset \B(\H)$ of a unital $*$-algebra $\As$ inside a semi-f\/inite von Neumann algebra
$\Ns\subset\B(\H)$ acting on the Hilbert space $\H$,
\item
a (non-necessarily bounded) self-adjoint operator $D$ on the Hilbert space $\H$ such that
\begin{itemize}\itemsep=0pt
\item
the domain $\dom(D)\subset \H$ of $D$ is invariant under all the elements $\pi(x)\in \pi(\As)$,
\item
the operators $[D,\pi(x)]_-$ def\/ined on $\dom(D)$ can be extended to bounded operators in the von Neumann algebra $\Ns$,
\item
for all $\mu\notin \Sp(D)$, the resolvent $(D-\mu I )^{-1}$ is a $\tau$-compact operator in $\Ns$ i.e.~it is in the norm closure of the ideal generated by all the projections $p=p^2=p^*\in \Ns$ with $\tau(p)<+\infty$.
\end{itemize}
\end{itemize}
\end{Definition}

\begin{Definition}\label{def: mst}
A \emph{modular spectral triple} $(\As,\H_\omega,D)$ relative to a semi-f\/inite von Neumann algebra $\Ns$ and a faithful
$\alpha$-KMS-state $\omega$ on the $*$-algebra $\As$ is given by:
\begin{itemize}\itemsep=0pt
\item
a faithful representation of $\As$ in $\Ns\subset\B(\H_\omega)$ where $(\pi_\omega,\H_\omega,\xi_\omega)$ is the
GNS-representation of $(\As,\omega)$;
\item
a faithful normal semi-f\/inite weight $\phi$ on $\Ns$ whose modular automorphism group $\sigma^\phi$ is inner in $\Ns$ and such that
$\sigma^\phi(\pi_\omega(x))=\pi_\omega(\alpha(x))$ for all $x\in \As$;
\item
$\phi$ restrict to a faithful semi-f\/inite trace $\tau:=\phi|_{\Ns^{\sigma^\phi}}$ on the f\/ixed point algebra $\Ns^{\sigma^\phi}\subset \Ns$;
\item
a (non-necessarily bounded) self-adjoint operator $D$ on $\H_\omega$ such that:
\begin{itemize}\itemsep=0pt
\item
$\dom(D)\subset \H_\omega$ is invariant under all the operators $\pi_\omega(x)$, $x\in\As$,
\item
for all $x\in \As$, $[D,\pi_\omega(x)]_-$ extends to a bounded operator in $\Ns$,
\item
for $\mu$ in the resolvent set of $D$, for all $f\in\pi_\omega(\As)^{\sigma^\phi}$, $f(D-\mu I)^{-1}$ is a $\tau$-compact operator relative to the semi-f\/inite trace $\tau$ on $\N^{\sigma^\phi}$.
\end{itemize}
\end{itemize}
\end{Definition}

Note that the def\/inition of modular spectral triple above has been formalized on the base of examples where the dynamical system
$(\As,\alpha)$ is periodic. Anyway, apart from the (of course essential) condition on the resolvent of $D$ (that allows to introduce semi-f\/inite Fredholm modules and index theory), most of the ingredients in the def\/inition are, in light of Connes--Takesaki duali\-ty and the properties of the Falcone--Takesaki non-commutative f\/low of weights in
Theorem~\ref{th: ct}, essentially canonical constructions in the modular theory of any von Neumann algebra equipped with a normal faithful semi-f\/inite weight! A f\/irst result going in this direction (Theorem~\ref{th: cpr}) will be provided at the end of
Subsection~\ref{sec: construction}.

\subsection{Modular theory and Antonescu--Christensen AF spectral triples} \label{sec: ac}

Let $\As$ be an AF $C^*$-algebra. Consider again a f\/iltration of $\As$ by unital inclusions of f\/inite dimensional $C^*$-algebras
$\As_0\subset \As_1\subset \cdots \subset \As_n \subset \As_{n+1}\subset \cdots \subset \As$ acting on the Hilbert space~$\H$ and
let $\xi \in \H$ be a cyclic and separating vector for the von Neumann algebra $\Rs:=\As''$.
We denote by $\Delta_\xi$ and $J_\xi$ the modular operator and the (conjugate-linear) modular conjugation operators relative to the pair $(\Rs,\xi)$ and by $(\sigma^\xi_t)_{t\in \RR}$ the corresponding one-paramenter group of modular automorphisms of $\Rs$.

For any $x\in \Rs$, def\/ine $\pi_n(x)\in\As_n$ by $\pi_n(x)\xi=P_n x\xi$, see e.g.~R.~Longo~\cite{L}.
\begin{Proposition}\label{pr: tak}
The following conditions are equivalent:
\begin{itemize}\itemsep=0pt
\item[a)]
$\pi_n: \Rs \to  \As_n$ is a $\ip{\xi}{\cdot\xi}$-invariant conditional expectation.
\item[b)]
$\sigma^\xi_t(\As_n)=\As_n$, $t\in \RR$.
\end{itemize}
Furthermore, $\Delta^{it}_\xi P_n=P_n\Delta^{it}_\xi$, $t\in \RR$ and $J_\xi P_n=P_n J_\xi$.
\end{Proposition}
\begin{proof}
Immediate from Takesaki's Theorem~\ref{th: cex}.
\end{proof}

\begin{Proposition}
If the filtration of the AF-algebra is modularly stable, for any choice of the sequence $(\alpha_n)$ that satisfies the
Antonescu--Christensen conditions $($as explained at the end of Section~$\ref{sec: st-ac})$, the corresponding Dirac operator is a modular invariant, i.e.:
\begin{equation*}
\Delta^{it}_\xi D \Delta^{-it}_\xi=D, \qquad \forall\,  t\in \RR, \qquad \text{and furthermore}, \qquad
J_\xi D=DJ_\xi.
\end{equation*}
\end{Proposition}
\begin{proof}
From Proposition~\ref{pr: tak} both $\Delta^{it}_\xi$ and $J_\xi$ commute with the projections $P_n$ and thus with the projection $E_n:=P_n-P_{n-1}$. \end{proof}

Therefore, in the non-tracial case, spectral triples associated to f\/iltrations that are stable under the modular group have a non-trivial group of automorphisms

We now provide some explicit example of AF-algebras whose f\/iltration is stable under a~non-trivial modular group.

\begin{Example}[Powers factors]
For every $n\in \NN_0$, let $\As_n:=\MM_2(\CC)\otimes \cdots \otimes \MM_2(\CC)$ be the tensor product of $n$ copies of
$\MM_2(\CC)$ and $\As_0:=\CC$.
Def\/ine by
$\iota_n: x \mapsto x\otimes 1_{\MM_2(\CC)}$ the usual unital inclusion $\iota_n: \As_n\to \As_{n+1}$.
For every $n\in \NN$, let $\phi_n:\MM_2(\CC)\to \CC$ be a faithful state and consider the (automatically faithful)
inf\/inite tensor product state
$\omega:= \phi_1\otimes \cdots \otimes \phi_n\otimes \cdots$ on the inductive limit $C^*$-algebra
$\As :=\otimes_{j=1}^\infty\MM_2(\CC)$.
Let $\Ms$ be the von Neumann algebra obtained as the weak closure of $\As$ in the GNS-representation of $\omega$.
We continue to denote with the same symbol the (automatically faithful)
normal extension of $\omega$ to $\Ms$.
If $\sigma^\omega$ denotes the modular group of $\Ms$, we have that
$\sigma^{\omega}_t(x_1\otimes\cdots\otimes x_n\otimes1\otimes\cdots)=\sigma^{\phi_1}_t(x_1)\otimes \cdots \otimes \sigma^{\phi_n}_t(x_n)\otimes1\otimes\cdots$.
The choice of
\begin{equation*}
\phi_j(x):=\text{tr}
\begin{bmatrix}
\lambda & 0 \\
0 & 1-\lambda
\end{bmatrix} x, \qquad j=1,\dots, n, \qquad \lambda\in ]0,1/2[,
\end{equation*}
gives rise to the so called Powers factor that is a factor of type III$_\mu$ with $\mu:=\lambda/(1-\lambda)$.
\end{Example}

\begin{Corollary}
The Antonescu--Christensen Dirac operators associated to the natural filtration of the Powers factors are modular invariant.
\end{Corollary}

As already implicit in C.~Antonescu, E.~Christensen's work~\cite[Theorem~2.1]{AC}, if we are willing to relax the condition on the compactness of the resolvent of the Dirac operator, the basic construction of the spectral triple can be carried on with respect to an arbitrary f\/iltration $\As_0\subset \As_1\subset \cdots \subset \As_n\subset \As_{n+1}\subset \cdots \subset \As$ of a $C^*$-algebra $\As$ without the assumption of the f\/inite dimensionality of the subalgebras $\As_n$.
This kind of construction seems to be related to the semi-f\/inite spectral triples developed by J.~Aastrup, J.~Grimstrup, R.~Nest in loop quantum gravity~\cite{AGN1,AGN2,AGN3,AGN4} (see also the discussion in Subsection~\ref{sec: otherqg}).

All this seems to suggest an interesting connection between spectral triples and modular theory: given a $\beta$-KMS state
$\omega$ over a $C^*$-algebra $\As$, for any (countable) f\/iltration of $\As$ with modular invariant subalgebras there are naturally associated families of modular invariant (semi-f\/inite) spectral triples obtained via (generalized) Antonescu--Christensen construction.

Since, for modular f\/iltrations of an AF $C^*$-algebra, the modular generator $K_\omega$ and the Dirac operators of Antonescu--Christensen commute, it is natural to ask if there are sitations where it is possible to assume $D=K_\omega$ or a proportionality between them.
Of course, it is already clear from the def\/inition that every Antonescu--Christensen Dirac operator has a positive spectrum
(since by construction $\alpha_n>0$), and hence the previous question should be interpreted in a~``loose way'' allowing some freedom for some (signif\/icant) alteration of the constructions.

In this respect it is worth to note the following elementary ``spectral symmetrization procedure'' on spectral triples: let $(\As,\H,D)$ be a ``Dirac ket'' spectral triple and consider $(\As,\H',D')$ its ``Dirac bra'' spectral triple, where $\H'$ denotes the Hilbert space dual of $\H$, the operator $D'$ is given by $D':=\Lambda\circ D\circ \Lambda^{-1}$, with $\Lambda:\H\to\H'$ the usual conjugate-linear Riesz isomorphism, and the $*$-algebra $\As$ is represented on $\H'$ via the faithful representation
$\pi'(x):=\Lambda\circ\pi(x)\circ \Lambda^{-1}$, for all $x\in \As$.

On the Hilbert space $\H\oplus\H'$, equipped with the direct sum $\pi\oplus \pi'$ of the faithful representations of
$\As$ on $\H$ and $\H'$, consider the self-adjoint operator $K:=D\oplus (-D')$. The operator $K$ has a symmetric spectrum very much resembling the situation typical of modular generators.
Furthemore it is possible to introduce a conjugate-linear operator $J:=\Psi\circ(\Lambda\oplus\Lambda^{-1})$, where
$\Psi:\H'\oplus\H\to\H\oplus \H'$ is the ``f\/lip operator'' $\Psi(\xi,\eta):=(\eta,\xi)$, for $\xi\in \H'$ and $\eta\in \H$.

A more interesting situation is provided via the tensor product construction. Given the same spectral triple $(\As,\H,D)$, consider the tensor product Hilbert space $\H\otimes\H'$, with its conjugate-linear ``f\/lip operator'' $J$ def\/ined on homogeneous tensors by
$J(\xi\otimes\eta):=\Lambda^{-1}(\eta)\otimes\Lambda(\xi)$, carrying the two commuting representations of the algebra $\As$ via
$\pi(x)(\xi\otimes\eta):=(\pi(x)\xi)\otimes \eta$ and $\pi'(x)(\xi\otimes \eta):=\xi\otimes(\pi'(x)\eta)=J\pi(x)J(\xi\otimes\eta)$, for all $x\in \As$, $\xi\in \H$ and $\eta\in \H'$. We can def\/ine a new Dirac operator $K:=D\otimes I - I\otimes D'$ on $\H\otimes\H'$ (remember that $D':=\Lambda\circ D\circ\Lambda^{-1}$) obtaining a new spectral triple $(\As,\H\otimes\H', K)$.
Even more general situations should be obtained via J.-L.~Sauvageot's relative tensor product (tensor product of correspondences).

Such a construction is not a completely futile exercise since it is typical of semi-f\/inite von Neumann algebras in their standard form and it will come handy in Subsection~\ref{sec: otherqg}.

As an elementary exemplif\/ication, we brief\/ly discuss here the case arising from the Cuntz algebra, somehow related to the Remark~2.2 in~\cite{AC}, and that has been already fully treated by A.~Carey, J.~Phillips, A.~Rennie~\cite{CPR3}.

Let $\O_d$ be the Cuntz algebra endowed with the faithful normal state $\omega=\tau\circ m$, where
$\tau$ is the canonical trace on the ``gauge invariant'' UHF-subalgebra $\O_d^0\subset \O_d$ and $m: \O_d\to \O_d^0$ is the conditional expectation obtained by averaging over the gauge group $\TT$.
We consider the natural $\ZZ$-grading of $\O_d$ and, for each $k \in \ZZ$, the projection $E_k$ from $\H_\omega$ (the GNS-space of $\omega$) onto the closed subspace $\cj{\pi_\omega(\O_d^k)\xi_\omega}$.

The modular one-parameter unitary group associated to the normal extension of $\omega$ to the von Neumann algebra $\pi_\omega(\O_d)''$, which is a factor of type III$_{1/d}$, is then
\begin{equation*}
t\mapsto \Delta_\omega^{it}=\sum_{k\in \ZZ} e^{-it\log(d) k} E_k.
\end{equation*}

Here the operator $K_\omega:=\log\Delta_\omega = - \log(d) \sum_{k\in \ZZ}  k E_k$
plays a role that is somehow analogous to the Dirac operator constructed by Antonescu--Christensen, apart from the requirement of compactness of the resolvent that cannot be satisf\/ied due to the inf\/inite dimensionality of its eigenspaces.
In particular, it is not dif\/f\/icult to check that all the commutators of $K_\omega$ with elements $\pi_\omega(a)$ with $a\in\O_d^k$ are bounded. For all $a\in \O_d^k$, for all $b\in \O_d$, we have that
$\pi_\omega(\sigma_t^\omega (a)b)\xi_\omega
=\Delta_\omega^{it}\pi_\omega(a\sigma^\omega_{-t}(b))\Delta_\omega^{-it}\xi_\omega
=\Delta_\omega^{it}\pi_\omega(a)\pi_\omega(\sigma^\omega_{-t}(b))\xi_\omega
=e^{-ikt\log(d)}\pi_\omega(a)\pi_\omega(b)\xi_\omega$,
since $\pi_\omega(a)\xi_\omega$ with $a\in \O_d^k$ is eigenvector of $\Delta_\omega$ with
eigenvalue $e^{-k\log(d)}$, and taking the derivative with respect to $t$ at $t=0$ we obtain $[K_\omega,\pi_\omega(a)]\pi_\omega(b)\xi_\omega=\pi_\omega(a)\pi_\omega(b)\xi_\omega$ and hence
$[K_\omega,\pi_\omega(a)]$ is bounded on the dense set $\pi(\O_d)\xi_\omega$.

Aside, we take the opportunity to mention a very natural question (that will also reappear later at the end of
Section~\ref{sec: meaning}).
Using f\/iltration techniques G.~Cornelissen, M.~Marcolli, K.~Reihani, A.~Vdovina~\cite{CMRV} have constructed some example of spectral triples over some particular Cuntz--Krieger algebra represented as a crossed product.
Is there a natural ``induction'' process for spectral triples?
Crossed products play a fundamental role in the structure of factors via Connes--Takesaki duality Theorem~\ref{th: ct} and it is natural to look for a theory of crossed products for spectral triples and in particular for modular spectral triples.

Finally, as a last remark in this section, we would like to spend a few general words on the interesting problems related to the mutual interplay between Atiyah--Singer index theory and modular theory. A.~Connes non-commutative geometry is deeply rooted in classical index theory. On the other side A.~Carey and collaborators have given for the f\/irst time an interpretation in terms of index of modular theory. Natural questions arise on the mutual interaction between the two: in most of the examples considered, modular spectral triples are constructed via a~``modular Dirac operator'' that is proportional to the modular generator and of course one would like to know if there are canonical ways to obtain a modular spectral triple out of a usual one and the other way around (this will be of the utmost importance for what we will say later on the reconstruction of space-time).
Very similar problems of ``transition'' between dif\/ferent forms of spectral geometry have been considered already, but only motivated by the dif\/ferential geometrical side of the matter, by J.~Fr\"ohlich and collaborators (see their discussion on the passage between $N=1$ and $N=(1,1)$ geometries or from spectral triples for conf\/iguration and phase-space~\cite{FGR3,FGR4}).
One is naturally led to conjecture that the already available incarnations of index theory may f\/ind their place in a wider landscape (that deserves to named geometry as well) where several spectral sets of data can be imposed. Spectral triples and modular spectral triples are probably just the f\/irst examples of this wider arena.

\subsection{Modular non-commutative geometry in physics}

Despite the fact that most, of the literature on non-commutative geometry is actually heavily motivated or directly inspired by physics (Heisenberg quantum mechanics, standard model, renormalization in perturbative quantum f\/ield theory, deformation quantization, just to mention a few) and the strong interest shared by theoretical physics for this mathematical subject, when compared to the outstanding structural achievements of Tomita--Takesaki modular theory in quantum statistical mechanics and algebraic quantum f\/ield theory, the fundamental relevance of non-commutative geometry for the foundations of physics looks still quite weak and disputable.
In this subsection rather than discussing the vast panorama of applications of non-commutative geometry to physics and model building (see the book by A.~Connes, M.~Marcolli~\cite{CM2} for a~recent very complete coverage of the physics applications of non-commutative geometry and, for a really pedestrian list of references, our companion survey paper~\cite{BCL6}), we proceed to describe the very few available instances and hints of a direct applicability of modular non-commutative ideas (such as semi-f\/inite and modular spectral triples, phase-spaces etc.) to physics.

In the investigation of models of supersymmetric quantum f\/ield theories A.~Jaf\/fe, A.~Les\-niews\-ki, K.~Osterwalder~\cite{JLO1,JLO2}
introduced $\theta$-summable Fredholm modules induced by KMS-functionals and D.~Kastler~\cite{K1} generalized the construction to supersymetric KMS-functionals (i.e.~graded KMS-states over a $\ZZ_2$-graded $C^*$-algebra with respect to a one-parameter group of automorphisms whose generator is the square of a ``supercharge'' operator). This seems to be in line with the conjectured deep structural relations between cyclic cohomology, supersymmetry and modular theory that, as reported by D.~Kastler~\cite{K1}, A.~Connes had already considered in his work on entire cyclic cohomology. Despite the signif\/icant interest attracted by this theme (see for example A.~Jaf\/fe, O.~Stoytchev~\cite{J,JS}), the construction of model independent supersymmetric KMS-states on algebras of observables of quantum f\/ields (say in algebraic quantum f\/ield theory) proved to be more elusive than expected and only with the recent works by D.~Buchholz, H.~Grundling~\cite{BGr1,BGr2} the road has been opened for the construction of supersymmetric cyclic cocycles and spectral triples in algebraic quantum f\/ield theory by S.~Carpi, R.~Hillier, Y.~Kawahigashi, R.~Longo~\cite{CHKL}.

R.~Longo~\cite{L3} has worked out the existence of deep relationships between supersymmetric cyclic cocycles obtained by
super-KMS-functionals and the theory of superselection sectors in algebraic quantum f\/ield theory (as references for superselection theory see for example D.~Kastler~\cite{K2} and R.~Haag~\cite[Chapter~IV]{H}).

Very deep and captivating motivations from physics permeate all of the works by
J.~Fr\"ohlich, O.~Grandjean, A.~Recknagel~\cite{FGR3,FGR4}, where several axiomatizations of non-commutative structures have been undertaken always with a view to physics. In particular, in the pre\-vious Section~\ref{sec: other}, we already described their def\/initions of non-commutative manifold and of non-commutative phase-space, that is especially important in our context, being the f\/irst non-commutative geometry making full-use of modular theory.

The already mentioned works by A.~Carey, J.~Phillips, A.~Rennie, F.~Sukochev and collaborators (see~\cite{CPR4} for a review) provide
def\/initions of semi-f\/inite and modular spectral triples with examples and techniques that, although mathematical in character, are without any doubt strongly motivated by the needs from physics (e.g.~A.~Jaf\/fe, A.~Lesniewski, K.~Osterwalder's cyclic cocycles) and provide vast opportunities for applications.

In particular, as we said, semi-f\/inite spectral triples already found their natural way in loop quantum gravity via the interesting works of J.~Aastrup, J.~Grimstrup, M.~Paschke, R.~Nest~\cite{AG1,AG2,AG3,AGN1,AGN2,AGN3,AGN4, AGNP, AGP,AGP2}.

Among the few approaches to a quantum theory of gravity via non-commutative geomet\-ry (see the companion
review~\cite[Section~5.5.1]{BCL6} for more complete references) A.~Connes, M.~Marcolli~\cite{CM2} make a systematic use of modular theory. They establish an extremely intriguing parallel between statistical mechanics methods applied to the study of number theory and a~tentative theory of quantum gravity. Using a ``cooling procedure'', introduced with C.~Consani in~\cite{CCM}, they examine the algebras of observables at dif\/ferent physical temperatures and they plan to recover macroscopic geometries via symmetry breaking.

The A.~Connes, M.~Marcolli's strategy, to recover non-commutative geometries out of symmetry breaking from a modular group acting on an algebra of observables for quantum gravity (obtained as ``convolution algebra'' of a $2$-category of kinematical quantum geometries), is further expanded in the remarkable recent paper by D.~Denicola, M.~Marcolli, A.-Z.~al~Yasri~\cite{DMaY} that is also introducing important constructions of topological spin networks and foams in view of direct applications of the formalism to loop quantum gravity.

\looseness=1
Very interesting speculations on the possible interplay between Tomita--Takesaki modular theory and non-commutative geometry are contained in M.~Paschke, R.~Verch~\cite[Section~6]{PV1} and suggestions on how the modular f\/low might be related to some aspects of (non-commutative) geometry have been conjectured by S.~Lord~\cite[Section~VII.3]{Lo}.
That non-commutative geometries (in the sense of A.~Connes' spectral triples or some variant of them) describing space-time might be recoverable from Tomita--Takesaki modular theory had been considered by R.~Longo~\cite{L2} and it has been the main theme of our investigation~\cite{B1,B3,BCL1,BCL6} for quite some time.

\section{Perspectives on modular algebraic quantum gravity}\label{sec: maqg}

Following~\cite{BCL1},
we propose\footnote{Bertozzini~P., Conti~R., Lewkeeratiyutkul~W., Modular algebraic quantum gravity, work in progress.} a ``thermal'' reconstruction of ``quantum realities'' (quantum space-time-matter), via Tomita--Takesaki modular theory, starting from suitable ``event states'' on ``categories'' of abstract operator algebras describing ``partial physical observables''.

The fundamental input of the project is the recognition that Tomita--Takesaki modular theo\-ry (the ``heart'' of equilibrium quantum statistical mechanics) can be reinterpreted as a way to associate non-commutative spectral geometries (axiomatically similar to A.~Connes' spectral triples) to appropriate states over the algebras of observables of a physical system.

In this way, for every ``observer'' (specif\/ied by an algebra of ``partial observables''), a dif\/ferent quantum geometry naturally emerges, induced by each possible ``covariant kinematic'' (specif\/ied by an ``event state''). These ``virtual realities'' interrelate with each other via a ``categorical covariance principle'' replacing the usual dif\/feomorphisms group of general relativity.

In our opinion this provides a new solid approach to the formulation of an algebraic (modular) theory of non-perturbative quantum gravity, and to the foundations of quantum physics, in which (quantum) space-time is reconstructed a posteriori\footnote{Our point of view, as it will appear in due course, is quite favorable to background independent approaches to quantum gravity as in the tradition of research in general relativity. The reader is warned that such a choice is not at all mainstream in the current research panorama in quantum gravity. For a very critical recent assessment of the interplay between quantum gravity and non-commutative geometry we suggest the recent review by J.~Gracia-Bond\'{\i}a~\cite{G-B}.}.

\subsection{Construction of modular spectral geometries } \label{sec: construction}

Building on our previous research program on ``modular spectral-triples in non-commutative geometry and physics''~\cite{B1,BCL1},  already announced in~\cite[Section~5.5.2]{B2,BCL6} and with more details in~\cite{B3}, we suggest here how to make use of
Tomita--Takesaki modular theory of operator algebras to associate non-commutative geometrical objects (only formally similar to A.~Connes' spectral-triples) to suitable states over involutive algebras (say for now $C^*$-algebras).

In the same direction we also stress the close connection of these ``spectral geometries'' to the modular spectral triples
introduced by A.~Carey, J.~Phillips, A.~Rennie, F.~Sukochev that we already described in Section~\ref{sec: mst}.
We proceed in steps.
\begin{itemize}\itemsep=0pt
\item
Let $\omega$ be a faithful $\beta$-KMS-state over the $C^*$-algebra $\As$.
By def\/inition there exists a~unique one-parameter group of $*$-automorphism $t\mapsto\sigma^\omega_t$ that satisf\/ies the
KMS-condition at inverse temperature $\beta$ for the state $\omega$.
\item
We consider the GNS-representation $(\pi_\omega,\H_\omega,\xi_\omega)$ induced by the state $\omega$ and note that the vector
$\xi_\omega$ is a cyclic and separating vector for the von Neumann algebra $\pi_\omega(\As)''\subset\B(\H_\omega)$.
This follows from O.~Bratteli, D.~Robinson~\cite[Vol.~2, Corollary~5.3.9]{BR}.
\item
Since $\xi_\omega$ is cyclic and separating for $\pi_\omega(\As)''$, we are entitled to apply Tomita--Takesaki theorem obtaining a unique one-parameter unitary group $t\mapsto \Delta_\omega^{it}$ on the Hilbert space $\H_\omega$ that spatially implements the modular group on the von Neumann algebra $\pi_\omega(\As)''$:
\begin{equation*}
\pi_\omega(\sigma^\omega_t(x))=\Delta^{it}_\omega \pi_\omega(x)\Delta^{-it}_\omega, \qquad \forall\, t\in \RR.
\end{equation*}
Again we denote by $K_\omega:=\log\Delta_\omega$ the modular generator and by $J_\omega$ the conjugate-linear involution, determined by $\xi_\omega$ on $\H_\omega$, that is spatially implementing a conjugate-linear isomorphism between
$\pi_\omega(\As)''$ and its commutant $\pi_\omega(\As)'$.

Note that the original $C^*$-algebra $\As$ is modularly stable i.e.~$\sigma^\omega_t(\As)=\As$ and that the restriction to $\As$ of the modular one-parameter automorphism group $t\mapsto\sigma^\omega_t$ of $\pi_\omega(\As)''$ coincides with the unique one-parameter group satisfying the
$\beta$-KMS condition on $\As$.
\item
Def\/ine $\As_\omega:=\{x\in \As \ | \ [K_\omega,\pi_\omega(x)]_- \in \pi_\omega(\As)''\}$ (where we mean here that the operator
$[K_\omega,\pi_\omega(x)]_-$ extends to a bounded operator in $\pi_\omega(\As)''$)  and note, by a direct application of
O.~Bratteli, D.~Robinson~\cite[Vol.~1, Theorem~3.2.61]{BR}, that $\pi_\omega(\As_\omega)\xi_\omega$ is a core for the operator $K_\omega$.
Furthermore, $\As_\omega$ is a $*$-algebra that is dense in $\As$.
\item
By Tomita--Takesaki theorem, for all $x,y\in \As_\omega$ we have $[[K_\omega,\pi_\omega(x)]_-,J_\omega\pi_\omega(y)J_\omega]_-=0$
that is very much resembling the f\/irst order condition in the def\/inition of A.~Connes' real spectral triples.
Clearly the other key condition in the def\/inition of real structures in spectral triples
$[\pi_\omega(x),J_\omega\pi_\omega(y^*)J_\omega]_-=0$ is always true for all $x,y\in \As_\omega$.

Furthermore, from the def\/inition of $\As_\omega$ we necessarily have $[K_\omega,\pi_\omega(x)]_-\in \pi(\As)''$, for all
$x\in \As_\omega$, that resembles one of the key ingredients in Def\/initions~\ref{def: s-st} and~\ref{def: mst} of modular spectral triples relative to a von Neumann algebra, although here the von Neumann algebra $\pi_\omega(\As)''$ is not necessarily
semi-f\/inite.
\end{itemize}

\begin{Definition}
Given a $\beta$-KMS state $\omega$ over the $C^*$-algebra $\As$, we def\/ine the \emph{modular spectral geometry} associated to the pair $(\As,\omega)$, to be the quintuple
\begin{equation*}
(\As_\omega,\H_\omega,\xi_\omega,K_\omega,J_\omega),
\end{equation*}
where the meaning of the terms above has been discussed in the items preceding this def\/inition.
\end{Definition}
As we already noted, when $\Ns:=\pi_\omega(\As)''$ is a semi-f\/inite von Neumann algebra, taking $D:=K_\omega$, this structure seems strictly related to the notion of modular spectral triple introduced by  A.~Carey, J.~Phillips, A.~Rennie.

Anyway, despite the superf\/icial resemblance of the data $(\As_\omega,\H_\omega,\xi_\omega,K_\omega,J_\omega)$ for our modular spectral geometries to actual real spectral triples, we have to stress a few radical dif\/ferences:
\begin{itemize}\itemsep=0pt
\item
the modular generator $K_\omega$ has a spectrum $\Sp(K_\omega)$ that is a symmetric set under ref\/lection in $\RR$ and can often be continuous, a situation that reminds of ``propagation'' and that does not have much in common with the usual f\/irst-order elliptic Dirac operators that appear in the def\/inition of A.~Connes' spectral triples.
\item
there is no grading anticommuting with $K_\omega$; although a natural grading is clearly present via the spectral decomposition of
$K_\omega$ into a positive and negative component, such grading always commutes with the modular generator.
\item
the resolvent properties of $K_\omega$ do not seem to f\/it immediately with the requirements of index theory, although we expect that in the case of periodic modular f\/lows the index theory developed by A.~Carey, J.~Phillips, A.~Rennie, F.~Suchocev for modular spectral triples will apply.
\item
contrary to the situation typical of A.~Connes spectral triples, the $*$-algebra $\As_\omega$ is stable under the one-parameter group  generated by $K_\omega$ that coincides with $t\mapsto\sigma^\omega_t$.
\end{itemize}
Of course, in order to justify the attribute ``geometry'' to be attached to an algebraic gadget, some more substance has to be present if not via index theory and (co)homology (as this is the case) at least through explicit presence of natural structures related to dif\/ferentiability, integral calculus \dots\ and it turns out that in our situation we have a few tools available:
\begin{itemize}\itemsep=0pt
\item
As is the case for any dynamical system, there is an intrinsic notion of smoothness provided by the modularly stable f\/iltration
$\As_\omega^{\infty}\subset \cdots \subset \As_\omega^{n+1}\subset \As_\omega^{n}\subset \cdots \subset \As_\omega^{0}\subset \As$
of $*$-algebras given, for $r\in\NN\cup\{+\infty\}$, by
$\As_\omega^r:=\{x\in \As \ | \ [t\mapsto\sigma_t^\omega(x)]\in C^{r}(\RR;\As)\}$, with $C^r(\RR;\As)$ denoting the family of $\As$-valued $r$-times continuously dif\/ferentiable functions.

Via the faithful representation $\pi_\omega $ this f\/iltration becomes a f\/iltration of smooth operators
$\pi_\omega(\As)^{\infty}\subset \cdots \subset \pi_\omega(\As)^{n+1}\subset \pi_\omega(\As)^{n}\subset \cdots \subset \pi_\omega(\As)^{0}\subset \pi_\omega(\As)\subset \pi_\omega(\As)''$ on the Hilbert space $\H_\omega$, where
$\pi_\omega(\As)^r:=\{x\in \As \ | \ [t\mapsto\Delta^{it}_\omega\pi_\omega(x)\Delta_\omega^{-it}]\in C^{r}(\RR;\B(\H_\omega))\}$ and now $C^r(\RR;\B(\H_\omega))$ denotes the set of $r$-times continuously dif\/ferentiable functions with values in the normed space
$\B(\H_\omega)$.
\item
With the previous smooth structure in mind, the operator $K_\omega$ seems to satisfy a variant of A.~Connes' regularity condition:
$\pi_\omega(x),[K_\omega,\pi_\omega(x)]_-\in\cap_{m=1}^{+\infty}\dom(\delta_\omega^m)$, where we assume
$\delta_\omega^m(x):=[K_\omega,\pi_\omega(x)]_-$.
\item
There is already a perfectly natural notion of integration available via the $\beta$-KMS state $\omega$ so that we can def\/ine
$\int x \, \text{d}\omega:=\omega(x)$, for all $x\in \As$ and more generally, for all $x\in \pi_\omega(\As)''$,
$\int x \, \text{d}\omega:=\ip{\xi_\omega}{x\xi_\omega}$.

Apart from the now trivial fact that it is usual in non-commutative measure theory to def\/ine Radon measures on spaces via states and more generally weights (see for example M.~Takesaki~\cite[Vol.~II, Chapter IX]{T}), it is well-known (see for example
A.~Jaf\/fe~\cite{J}, J.~Fr\"ohlich, O.~Grandjean, A.~Recknagel~\cite[Section~5.1.3]{FGR3}, \cite[Def\/inition~2.1.3]{FGR4},
and also R.~Longo~\cite{L2,L3}) that in non-commutative integration theory for f\/inite and $\theta$-summable spectral triples, Dixmier traces can be substituted by functionals involving trace of the  ``heat kernel'' $\exp(-tD^2)$. Furthermore, for the general case of ``non-compact inf\/inite-dimensional spaces'', it has been already suggested (see again~\cite[Section~2.2.6]{FGR4},
\cite[Section~5.2.6]{FGR3} and especially A.~Jaf\/fe~\cite[Section~2]{J}) that $\beta$-KMS states should be used for the def\/inition of an integration theory.
\item
Note that from the above def\/inition, for the purpose of integration, the order-one operator $K_\omega$ plays a role similar to the Laplacian $D^2$ in the case of spectral triples.
\end{itemize}

Although we restricted our discussion here to the case of $\beta$-KMS states on a $C^*$-algebra, the construction of a modular spectral geometry associated to the pair $(\As,\omega)$ can be performed in the more general situation of faithful semi-f\/inite normal weights.

Making use of Connes--Takesaki duality and Falcone--Takesaki construction of the non-com\-mu\-ta\-ti\-ve f\/low of weights we obtain immediately  some important relation between general mo\-du\-lar spectral geometries and a semi-f\/inite version of them resulting in structures that are really very close to those of ``modular spectral triples'' by A.~Carey and collaborators and that deserve a careful study in order to identify their physical signif\/icance.
\begin{itemize}\itemsep=0pt
\item
Def\/ine $\M_\omega:=\pi_\omega(\As)''$ and note that $\M_\omega$ is canonically embedded in a semi-f\/inite von Neumann algebra
$\widetilde{\M}_\omega$, isomorphic to the crossed product $\M_\omega\rtimes_{\sigma^\omega}\RR$, and hence $\M_\omega$ is identif\/ied with the algebra of f\/ixed points $\widetilde{\M}_\omega^{\widehat{\sigma}^\omega}$ for the dual action
$s\mapsto\widehat{\sigma}^\omega_s$ on $\widetilde{\M}_\omega$.
\item
There is an operator-valued weight $\Xi_\omega$ from $\widetilde{\M}_\omega$ to $\M_\omega$ under which the weight $\omega$ get lifted to
$\widetilde{\M}_\omega$ as $\widetilde{\omega}:=\omega\circ\Xi_\omega$ and $\M_\omega$ inside $\widetilde{\M}_\omega$ is modularly stable under the modular group $t\mapsto\sigma^{\widetilde{\omega}}_t$ of $\widetilde{\omega}$ on
$\widetilde{\M}_\omega$.
\item
The modular group $t\mapsto\sigma^{\widetilde{\omega}}_t$ is inner in $\widetilde{\M}_\omega$ with generator $k_\omega$ af\/f\/iliated to $\widetilde{\M}_\omega$.
\item
There is a natural trace on $\widetilde{\M}_\omega$ def\/ined by
$\tau_\omega(z):=\widetilde{\omega}(\omega^{-1/2}z\omega^{-1/2})$, for all
$z\in \widetilde{\M}_\omega$, where $\omega^{-1}:=\exp(-k_\omega)$. Furthemore, the natural trace rescales under the action of
$s\mapsto\widehat{\sigma}^\omega_s$ i..e.~$\tau_\omega\circ \widehat{\sigma}^\omega_s=e^{-s}\tau$.
\end{itemize}

Much more concrete results can be obtained along the following lines.

Consider the von Neumann algebras
$\M^\omega:=\{x\in \M_\omega \ | \ \sigma^\omega_t(x)=x, \ \forall t\in \RR\}$,
usually called the centralizer of $\omega$, and
$\N_\omega:=\{\pi_\omega(x)\Delta_\omega^{it} \ | \ x\in \As, \ t\in \RR\}''
=\M_\omega \vee \{\Delta_\omega^{it} \ | \ t \in \RR\}''$.
We clearly have that $\M^\omega\subset \M_\omega\subset \N_\omega$.
Notice that $\M^\omega=\M_\omega \cap \{\Delta_\omega^{it} \ | \ t \in \RR\}'$ and therefore
$(\M^\omega)'=(\M_\omega)' \vee \{\Delta_\omega^{it} \ | \ t \in \RR\}''$.

All the previous algebras are acting on the Hilbert space $\H_\omega$ where $\M_\omega$ is in standard form.
Passing to the commutant von Neumann algebras we obtain $\N_\omega'\subset \M_\omega'\subset(\M^\omega)'$ and,
since $\gamma_\omega(\M_\omega)=\M_\omega'$, one has $\gamma_\omega((\M^\omega)')=\N_\omega$.
In particular, being (anti-isomorphic to) the commutant of a (semi-)f\/inite von Neumann algebra, $\N_\omega$ is
semi-f\/inite~\cite[Corollary V.2.23]{T}.

By \cite[Theorem~VIII.2.6]{T},
the state $\omega$ on $\M_\omega$ restricts, to a trace on $\M^\omega$.

Since $\M^\omega$ is modularly stable under $\sigma^\omega$, by Theorem~\ref{th: cex} there is a unique conditional expectation
$\Phi_\omega:\M_\omega\to \M^\omega$ such that $\omega|_{\M^\omega}=\omega\circ\Phi_\omega$ and via the conjugate-linear map $\gamma_\omega$ we obtain a unique conditional expectation $\Phi_{\gamma_\omega}: \M_\omega'\to\N_\omega'$ given,
for all $x\in \M_\omega'$, by
$\Phi_{\gamma_\omega}:x\mapsto \gamma_\omega\circ \Phi_\omega\circ\gamma_\omega(x)$.

Making use of~\cite[Theorem~IX.4.24]{T}, we can now associate to the conditional expectation
$\Phi_{\gamma_\omega}:\M_\omega'\to\N_\omega'$ a dual faithful semi-f\/inite normal operator valued weight
$\Theta_\omega:\N_{\omega+}\to\M_{\omega+}$.

Such an operator-valued weight can be used to lift the state $\omega$ on $\M_\omega$ to a faithful normal semi-f\/inite weight
$\phi_\omega$ on $\N_\omega$ such that $\phi_\omega:=\omega\circ \Theta_\omega$.

By Theorem~\ref{th: ovw} we have that
$\sigma^{\phi_\omega}_t(x)=\sigma^\omega_t(x)$ for all $x\in \M_\omega$ and for all $t\in \RR$.

Using the def\/inition of $\Theta_\omega$ and the properties of spatial derivatives (see Theorem~\ref{th: sp-der}),  it is straightforward to verify that the modular group induced by the weight
$\phi_\omega$ on the semi-f\/inite von Neumann algebra $\N_\omega$ is given, for all $x\in \N_\omega$, by
$\sigma^{\phi_\omega}(x)=\Delta_\omega^{it}x\Delta_\omega^{-it}$.
By~\cite[Theorem~VIII.3.14]{T} the map def\/ined for all $x\in \N_\omega$ by $\tau_\omega(x):=\phi_\omega(\Delta_\omega^{-1/2}x\Delta_\omega^{-1/2})$ is a faithful semi-f\/inite normal trace.

Making full use of the notation that we have just introduced above, we can now state the following result
on rather general grounds.
\begin{Theorem} \label{th: cpr}
Let $(\As_\omega,\H_\omega,\xi_\omega,K_\omega,J_\omega)$ be the modular spectral geometry associated to the pair
$(\As,\omega)$, where $\omega$ is an $\alpha$-KMS-state over the $C^*$-algebra $\As$. Suppose that $K_\omega$ has compact resolvent with respect to the canonical trace $\tau_\omega$ on the von Neumann algebra $\N_\omega$ and that the extended weight
$\phi_\omega:\N_{\omega+}\to [0,+\infty]$ is strictly semi-finite\footnote{By a strictly semi-f\/inite weight $\phi:\N_+\to[0,+\infty]$ we mean that a weight $\phi$ whose restriction to its centralizer is semi-f\/inite.}. The data $(\As_\omega,\H_\omega,K_\omega)$ canonically provide a modular spectral triple, relative to the von Neumann algebra $\N_\omega$ and to the $\alpha$-KMS-state $\omega$, according to A.~Carey, J.~Phillips, A.~Rennie Definition~{\rm \ref{def: mst}}.
\end{Theorem}

\begin{proof}
We already know that $\As_\omega:=\{x\in \As \ | \ [K_\omega,x]_-\in \M_\omega\}$ is a $*$-algebra that is $\alpha$-invariant inside the $C^*$-algebra $\As$. By hypothesis $\omega$ is an $\alpha$-KMS state on $\As$ and $\H_\omega$ is the Hilbert space of the
GNS-representation $(\pi_\omega,\H_\omega,\xi_\omega)$ of $(\As,\omega)$.

The GNS representation $\pi_\omega$ is a covariant representation for the one-parameter group $\alpha:t\mapsto \alpha_t$ of automorphisms of $\As$ that is implemented on $\H_\omega$ by the modular one-parameter group $t\mapsto e^{iK_\omega t}$
i.e.~$\pi_\omega(\alpha_t(x))=e^{iK_\omega t}\pi_\omega(x)e^{-iK_\omega t}$, for all $x\in\As$ and $t\in \RR$.

We have already seen from the def\/inition of $\N_\omega$ that the modular generator $K_\omega$ is af\/f\/iliated to~$\N_\omega$ and that actually induces an inner one-parameter automorphism group of $\N_\omega$ that coincides with the modular group of the semi-f\/inite normal faithful weight $\phi_\omega:=\omega\circ\Theta_\omega$, where $\Theta_\omega$ is the operator-valued weight from
$\N_\omega$ to $\M_\omega:=\pi_\omega(\As)''$ constructed above.
Since $\sigma^{\phi_\omega}|_{\M_\omega}=\sigma^\omega$, we have
$\sigma^{\phi_\omega}(\pi_\omega(x))=\sigma^\omega(\pi_\omega(x))=\pi_\omega(\alpha(x))$, for all $x\in \As$.

Since $\Delta_\omega^{-1/2}\in \N_\omega$, from the formula
$\tau_\omega(z):=\phi_\omega(\Delta_\omega^{-1/2}z\Delta_\omega^{-1/2})$, for all $z\in \N_\omega$,
for the canonical trace $\tau_\omega$ on $\N_\omega$, we obtain
$\phi_\omega(z)=\tau_\omega(\Delta_\omega^{1/2}z\Delta_\omega^{1/2})$, for all $z\in \N_\omega$.
By assumption, $\phi_\omega|_{\N^{\sigma^{\phi_\omega}}}$ is a semi-f\/inite trace.

The operators $[K_\omega,\pi_\omega(x)]_-$, for all $x\in \As_\omega$ extend to operators in
$\M_\omega\subset \N_\omega$. Since $\pi(\As_\omega)\xi_\omega$ is a core for the operator $K_\omega$ that is invariant for all the bounded operators $\pi_\omega(x)$ with $x\in \As_\omega$, we have that $\dom(K_\omega)$ is invariant under all $\pi_\omega(x)$, for $x\in \As_\omega$. Finally the hypothesis on the $\tau_\omega$-compactness of the resolvent of $K_\omega$ concludes the proof.
\end{proof}
\begin{Remark}
Clearly, in the reverse direction, if $(\As,\H_\omega,D)$ is a modular spectral triple relative to a semi-f\/inite von Neumann algebra
$\N\subset \B(\H_\omega)$ and an $\alpha$-KMS state $\omega$ on $\As$ whose Dirac operator $D$ concides with the modular generator $K_\omega$ of the modular one-parameter group $t\mapsto \Delta_\omega^{it}$ on $\H_\omega$, the uniquely determined data $(\As_\omega,\H_\omega,\xi_\omega,D,J_\omega)$ provides the modular spectral geometry associated to the pair
$(\As,\alpha)$.

Since, as far as we know, all the interesting examples of modular spectral triples available for now are equipped with a Dirac operator that is proportional to the modular generator $K_\omega$, the observation above says that modular spectral triples are ``essentially'' very specif\/ic modular spectral geometries.

One might also wonder to which extent
the property of $\tau_\omega$-compactness of the resolvent of~$K_\omega$ and strict semi-f\/initeness of $\widetilde{\omega}$ might be features related to the specif\/ic examples that are now reasonably well understood from the point of view of modular index theory i.e.~the case of periodic modular groups.
\end{Remark}

From Connes--Takesaki duality theorem it is clear that the initial physical information coming from the pair $(\As,\omega)$ and codif\/ied into the von Neumann dynamical system $(\M_\omega, \sigma^\omega)$ is equivalent to the information codif\/ied via the semi-f\/inite von Neumann dynamical system $(\N_\omega,\widehat{\sigma}^\omega)$ and hence there will be a way to relate the modular spectral geometry associated to the pair $(\As,\omega)$ and the one associated to $(\N_\omega,\widetilde{\omega})$ and we can take advantage of the semi-f\/inite case.

We are currently examining how the categorical properties of Tomita--Takesaki modular theory will translate into functorial properties for the previous constructions of modular spectral geometries. As we will say in Subsection~\ref{sec: categ} we think that this will be a key point for the study of covariance in quantum gravity, but this will be done elsewhere.

\subsection{Physical meaning of modular spectral geometries}\label{sec: meaning}

We now deal with the possible physical interpretation of the ``spectral geometries'' that have been constructed in the previous section. In our opinion this is probably the most signif\/icant original message conveyed by this manuscript: what is essentially the
well-known modular theory of KMS-states, standard forms of von Neumann algebras and their already clear link with equilibrium quantum statistical mechanics will require some ``reinterpretation'' to f\/it in the covariant context.
In this case the mathematical formalism (KMS-states over $C^*$-algebras) is already in place, but a dif\/ferent interpretative scheme has to be constructed. We are aware that our proposal will encounter immediately  several important objections and so we will devote some time to address at least some of these issues.

Loosely speaking, the most surprising thing in Tomita--Takesaki theory is that the knowledge of certain states over the algebra of observables (that, in the usual interpretation of algebraic quantum mechanics, means knowledge of the information on the past preparation of a physical system) entails the complete knowledge of the dynamics of the system.

In a similar way, although in a completely dif\/ferent context, it is recognized that Einstein's equations are a recipe for extracting information on the space-time geometry (at least its metric) from the knowledge of the distribution of matter (the stress-energy tensor). It has also been conjectured (for example by S.~Doplicher~\cite{Do1,Do4}) that in a theory of quantum gravity the geometry will depend on the ``state'' of the system.
\begin{itemize}\itemsep=0pt
\item
Our main conjecture is that Tomita--Takesaki theorem plays a role of quantum Einstein equation, providing a way to associate to a suitable ``state'' of the physical system a dynamics, and that this dynamics can be interpreted as a certain
non-commutative geometry (although not necessarily or immediately at the level of space-time).
\end{itemize}

We are perfectly aware that, in the context of C.~Rovelli ``thermal time hypothesis'', the modular automorphisms group has been proposed as a way to recover a notion of ``classical time''~\cite{CR}.

Furthermore, both in the ``termal time hypothesis'' as well as in the more traditional usage of the modular f\/lux as the one-parameter group of time evolution of a quantum statistical system in equilibrium, the modular Hamiltonian $K_\omega$ generates only a
``one-dimensional'' group and it has usually been said that the knowledge of such a ``one-dimensional thing'' as the modular group (or equivalently the modular Hamiltonian) will not be suf\/f\/icient to gather information for the other (missing spatial) dimensions, let alone for ``quantum geometries'', since the parameter in the f\/lux is just a classical real number. We point out that:
\begin{itemize}\itemsep=0pt
\item
The modular generator $K_\omega$ is a self-adjoint operator and every self-adjoint operator on a Hilbert space (including for example the Dirac operator in the very same def\/inition of spectral triple!)\ is the generator of a one-parameter group of unitaries on the Hilbert space. For example, the fact that the Atiyah--Singer Dirac operator $D$ generates on the Hilbert space $\H$ of square-integrable sections of the spinor bundle of a spinorial Riemannian manifold $M$ a unitary one-parameter group of unitaries $t\mapsto \exp(itD)$ in a classical parameter $t\in \RR$, does not prevent one from extracting the full information on the manifold $M$ from the data codif\/ied in the spectral triple $(C^\infty(M),\H,D)$.
\item
Although we agree that C.~Rovelli ``thermal time hypothesis'' can still be a perfectly viable possibility to obtain a macroscopic classical time evolution in situations (such as generally covariant relativistic systems) where a notion of (classical or quantum) time might be absent from the basic theory, we stress that extracting a macroscopic time evolution from a statistical state is something that can be done even on a classical system with a non-pure statistical state and hence does not seem to require a non-commutative algebra of observables.

To the contrary, Tomita--Takesaki modular theory carries information on the deep quantum structure of the algebra of observables (in fact the modular f\/low becomes trivial in the commutative case!).

More precisely, one expects that obtaining a ``commutative geometry'', from the ``quantum geometry'' supposedly described by modular theory, will of course require some form of ``coarse graining'' for the very same reason that the modular geometry induced by the modular f\/low becomes trivial on commutative $C^*$-algebras and so the construction cannot be immediately applied to the limiting classical case. What looks suspicious, is that the thermal time hypothesis works well in the classical case too, even if the corresponding modular f\/low is trivial.
\item
The usual interpretation of KMS-states as equilibrium states in quantum statistical mechanics for the modular one-parameter group and hence the association of Tomita--Takesaki theory with a time evolution is based on ``non-relativistic quantum theory''.

In a quantum theory of gravity, or for generally covariant quantum systems, there are already a few tentative versions of ``relativistic quantum theory'', mainly in the Schr\"odinger picture (see for example
C.~Rovelli ``relativistic quantum mechanics''~\cite[Section~5.2]{Ro1}, K.~B\"ostrom ``event quantum theory''~\cite{Bos1,Bos2} and C.~Isham ``history projection quantum theory''~\cite{I1,I2,IL1,IL2,ILSS,Sav1,Sav2}), in which a Hilbert space formalism of states and observables still survives and where the dynamics is not described in terms of time evolution, but via operator contraints (Wheeler--DeWitt equations), propagators or decoherence functionals.

What we would like to suggest is that a ``fully covariant'' interpretation of the modular group might be possible, where the evolution parameter is a ``relativistic scalar'', moreover in such a context a ``covariant thermal \textit{space-time} hypothesis'' should be viable\footnote{By this we mean that both macroscopic time and space degrees of freedom should be recoverable from the knowledge of a statistical state for a covariant theory.}.
\end{itemize}

Since this last interpretative issue is crucial, we are trying to develop\footnote{Bertozzini~P., Algebraic formalism for Rovelli quantum theory, in preparation.} an ``event interpretation'' of the formalism of states and observables in algebraic quantum physics that is in line with the above mentioned C.~Isham's ``history projection operator theory'' and/or with C.~Rovelli's ``relativistic quantum theory'', to which we will constantly refer in the following.

Without entering here in the discussion of details, that will be fully addressed elsewhere, we note that the data in the specif\/ication of a modular spectral geometry $(\As_\omega,\H_\omega,K_\omega)$ look essentially similar to those in C.~Rovelli's relativistic quantum theory~\cite[Section~5.2]{Ro1} with:
\begin{itemize}\itemsep=0pt
\item
the $C^*$-algebra $\As=\As_\omega^-$ as the algebra of covariant partial observables of an observer,
\item
the modular generator $K_\omega$ as the covariant dynamical constraint,
\item
the f\/ixed point algebra $\As^{\sigma^\omega}$ with respect to the modular one-parameter group $t\mapsto \sigma^\omega_t$ as the algebra of complete covariant observables: $x\in \As^{\sigma^\omega} \iff [K_\omega,\pi_\omega(x)]_-=0$.
\end{itemize}
The parallel is unfortunately far from being perfect indeed:
\begin{itemize}\itemsep=0pt
\item
The Hilbert space $\H_\omega$ is ``too big'' to be identif\/ied with the space of kinematical states: the GNS representation
$\pi_\omega$ for the KMS-state $\omega$ is not irreducible (the commutant $\pi_\omega(\As)'$ is conjugate-isomorphic to
$\pi_\omega(\As)''$). Some ``kind of polarization'' in $\H_\omega$ is needed to recover the covariant kinematical Hilbert space.
\item
It is tempting to think that $\H_\omega$ might be related to C.~Rovelli's boundary Hilbert space.
\item
The vector state $\xi_\omega\in \H_\omega$ in this case will play the role of covariant vacuum.
\item
It is not immediately clear what kind of ``non-commutative geometry'' is described by the modular spectral geometries
$(\As_\omega,\H_\omega,\xi_\omega,K_\omega,J_\omega)$ reconstructed from Tomita--Takesaki modular theory, but they seem to be geometries related to the ``phase space'' of the physical system and actually the structures available in the case of modular spectral geometries are intriguingly close to those in the axioms of ``quantized phase space'' early proposed by
J.~Fr\"ohlich, O.~Grandjean, A.~Recknagel~\cite[Section~5.2.6]{FGR3}, \cite[Section~2.2.6]{FGR4}.

Note that by ``phase space'' here we do not mean the quantum version of the C.~Ro\-vel\-li's relativistic phase space $\Gamma$ (the space of solutions of the dynamical constraints), but rather the ``boundary space'' i.e.~the quantum version of the space of ``boundary data''~$\G$  introduced by C.~Rovelli in the context of relativistic Hamiltonian dynamics~\cite[Section~3.2.2]{Ro1}.
The Hilbert space of square-integrable functions on $\G$ would be identif\/iable with the Koopman--von Neumann Hilbert space of classical mechanics that is used in operator algebraic versions of non-relativistic quantum mechanics on phase-space.
\end{itemize}

In order to obtain information on the geometry of the ``relativistic conf\/iguration space'' of the physical system, unfortunately we have to face, in a covariant context, the same kind of problems already indicated by
J.~Fr\"ohich, O.~Grandjean, A.~Recknagel~\cite[Section~2]{FGR3}, \cite[Section~5.4]{FGR4} about the relation between non-commutative geometries on ``phase space'' built out of spectral data $(\As,\H_\As,K_\As)$ involving primarily the algebra of observables $\As$ and spectral data $(\Cs,\H_\Cs,D_\Cs)$ involving the algebras of coordinates $\Cs$ on a ``conf\/iguration space''.

As already observed by J.~Fr\"ohlich and collaborators~\cite[Section~2]{FGR3}, it is possible to expect the appearance of ``target duality'' phenomena i.e.~the fact that dif\/ferent geometries on ``conf\/iguration space'' might be compatible with the overall unique geometry on ``phase space''.

A very promising route to address this delicate circle of problems is using ideas developed by M.~Paschke~\cite{P3,P1} and
T.~Kopf, M.~Paschke~\cite{KP3} and consists of selecting, inside the algebra of observables $\As$, a certain subalgebra $\Cs$ of ``coordinates'' that satisf\/ies appropriate axioms of what they call ``scalar quantum mechanics''~\cite{P1,KP3}
(see also J.~Kot\r{u}lek~\cite{Kot1,Kot2}).

Among the possible criteria of selection of suitable subalgebras of coordinates, we are also particularly intrigued by the recent work on ``non-commutative Cartan subalgebras'' by R.~Exel~\cite{Ex} (see also the related important work by J.~Renault~\cite{Ren}): the algebra of observables $\As$ should be reconstructed from an ``algebra of coordinates'' via a kind of ``crossed product''.

To exploit this ideas, a full theory of crossed products for spectral triples (and more generally non-commutative spectral geometries) should be developed (some interesting results on crossed products for spectral triples by discrete groups have been obtained by G.~Cornelissen, M.~Marcolli, K.~Reihani, A.~Vdovina~\cite{CMRV}).
In the special case of modular spectral geometries, Connes--Takesaki duality should be particularly relevant in this context and work on this front is ongoing\footnote{Bertozzini~P., Conti~R., Lewkeeratiyutkul~W., Modular spectral triples, in preparation.}.

Finally we would like to stress that although here we did not yet address in any signif\/icant way the important problems related to the exploration of the physical meaning of the modular f\/lows $\mapsto \sigma^\omega_t$ and $t\mapsto\Delta_\omega^{it}$, some ideas about their possible relations with a renormalization f\/lux and the contruction of ``action functionals'' obtained by taking the modular generator $K_\omega$ in place of the usual Dirac operator in non-commutative geometry are under consideration.
We will return to these points in the future.

\subsection{Categories of modular spectral geometries}\label{sec: categ}

In the previous two sections we have been considering only $\beta$-KMS states $\omega$ over a $C^*$-al\-geb\-ra~$\As$; we have been using these data $(\As,\omega)$ to construct structures $(\As_\omega,\H_\omega,\xi_\omega,K_\omega,J_\omega)$ that have a distinct non-commutative geometric f\/lavour; we have put forward some tentative interpretation for these modular spectral geometries as geometries of ``phase space'' (in the sense of boundary space) of a physical system and we proposed some strategies to obtain from them non-commutative geometries on conf\/iguration space. Even after this last step, we are still very far from having in our hands something that might be called or compared with an actual (non-commutative) space-time geometry.

Of course we are well aware of the possibility that, at a fundamental level, only a very generalized notion of space-time (in this case phase space or conf\/iguration space), if one at all, might survive and that space-time as we know it might just be a macroscopic construct emerging in some coarse graining (this is for example the position of A.~Connes, M.~Marcolli~\cite{CM2} when they state that geometry emerges by symmetry breaking at low temperatures); but we think that the questions posed by the identif\/ication of degrees of freedom related to the spatio-temporal organization of the physical system should deserve a more careful analysis also because ``space-time'' in our current theories is actually hiding completely dif\/ferent and possibly unrelated properties: localization, separability, causality, covariance, (local) symmetry.

After all, one of the most breathtaking achievements of A.~Connes has been to show that the Einstein's program of geometrization of physics can be carried out in a clean way via non-commutative geometry \dots\ and it seems too early to leave this ``paradise regained'' at the level of a fundamental theory just for pleasing fashionable trends.

As a f\/irst step in the direction of clarif\/ication of the several roles played by space-time in fundamental physics, we turn to the axioms of algebraic quantum f\/ield theory in Araki--Haag--Kastler form~\cite{H,A}. In this setting, a quantum theory of f\/ields on Minkowski space-time $\MM$, is specif\/ied by the assignment of a net $\O\mapsto \As(\O)$ that associates unital $C^*$-subalgebras of a
given unital $C^*$-algebra $\As(\MM)$ (the ``quasi-local algebra of observables'') to every open region $\O$ in Minkowski space-time.
A minimal set of axioms comprise the f\/irst four in this list:
\begin{itemize}\itemsep=0pt
\item[--]
(isotony) $\O_1\subset \O_2$ implies $\As(\O_1)\subset \As(O_2)$,
\item[--]
(locality) if $\Os_1$ and $\O_2$ are space-like separated we have $[\As(\O_1),\As(\O_2)]_-=\{0\}$,
\item[--]
(covariance) there is a representation $\alpha:g\mapsto \alpha_g$ of  the proper orthochronous Poincar\'e group $\P_+^\uparrow$ of Minkowski space as group of $*$-automorphisms of the quasi-local $C^*$-algebra $\As(\MM)$ acting geometrically on the net,
i.e.~$\alpha_g(\As(\O))=\As(g\O)$, $\forall \, g\in \P_+^\uparrow$ and $\forall \O\subset \MM$,
\item[--]
(primitive causality) $\As(\O)=\As(\widetilde{\O})$, where $\widetilde{\O}$ denotes the causal completion of $\O$,
\item[--]
(vacuum state) there is a pure state $\omega$ on $\As$ that is invariant under the Poincar\'e group,
i.e.~$\omega\circ\alpha_g=\omega$ for all $g\in \P_+^\uparrow$, whose associated GNS-representation
$(\pi_\omega,\H_\omega,\xi_\omega)$ is covariant for the action of $\P_+^\uparrow$, i.e.~there is a unitary representation
$g\mapsto U_g$ such that,
 for all $x\in \As$ and $g\in \P_+^\uparrow$,
 $\pi_\omega(\alpha_g(x))= U_g\pi_\omega(x)U_g^{-1}$
and moreover $\xi_\omega$ is the unique (up to phase) $U$-invariant vector.

\item[--]
(positivity) the joint spectrum of the generators of the subgroup of translations $v\mapsto U_v$ (where $U_v$ is the unitary corresponding to the traslation by the vector $v\in \MM$) is contained in the closure of the forward light-cone.
\end{itemize}

Following the ideas of J.~Dimock~\cite{Dim1,Dim2}, the previous set of axioms has been signif\/icantly generalized in a
``functorial way'' f\/irst by S.~Hollands, R.~Wald~\cite{HW} and R.~Verch~\cite{V}, then by
R.~Brunetti, K.~Fredenhagen, R.~Verch~\cite{BFV} (see also  R.~Brunetti, R.~Porrmann, G.~Ruzzi~\cite{BPR}) in order to cover not only cases of f\/ields living on a f\/ixed curved space-time, but also generally covariant situations. In this case a quantum theory of f\/ields is assigned via:
\begin{itemize}\itemsep=0pt
\item[--]
(isotony+covariance) a covariant functor $\As:\O\mapsto\As(\O)$, $\As:\psi\mapsto\As_\psi$ from a category $\Mg$ of globally hyperbolic four-dimensional oriented and time-oriented Lorentzian manifolds (with orientations and causality preserving isometries as maps) to a category $\Ag$ of unital $C^*$-algebras (with unital $*$-homomorphisms),
\item[--]
(locality) whenever there is a couple of morphisms $\psi_1:\O_1\to\O$ and $\psi_2:\O_2\to\O$ such that $\psi_1(\O)$ and $\psi_2(\O)$ are causally separated in $\O$, we have commutativity of observables $[\As(\O_1),\As(\O_2)]_-=\{0\}$,
\item[--]
(primitive causality) whenever a morphism $\psi:\O_o\to\O$ is such that $\psi(\O_o)$ contains a~Cauchy surface for $\O$, we have that $\As_\phi:\As(\O_0)\to\As(\O)$ is an isomorphism.
\end{itemize}
Quantum f\/ields for such a locally covariant quantum f\/ield theory $\As$ are identif\/ied with natural transformations from $\Ds$ to $\As$,
where $\Ds:\Mg\to\Ag$ is the covariant functor that associates to manifolds their algebras of smooth test functions.

As far as we know, all the available examples of quantum f\/ield theories that satisfy the previous set of axioms are obtained as
\textit{free fields} via a second quantization functor (see for example the very clear and recommended exposition in the book by
C.~B\"ar, N.~Ginoux, F.~Pf\"af\/f\/le~\cite{BGP}) associating Weyl algebras to symplectic spaces induced by propagators of  normally hyperbolic operators on globally hyperbolic oriented and time oriented Lorentzian manifolds.

Returning to the axioms of algebraic quantum f\/ield theory, we note that following J.~Schwin\-ger, contrary to the case of ordinary quantum mechanics, space-time coordinates do not appear any more as operators and special covariance is restored downgrading coordinates to mere book-keeping devices for the algebras of observables in the axioms of isotony and locality\footnote{There are actually several approaches trying to ``redef\/ine'' operators associated to coordinates (in particular time) in an algebraic quantum f\/ield theory context. Among them we refer to R.~Brunetti, K.~Fredenhagen, M.~Hoge~\cite{BFr,BFH}.}.

Still the degrees of freedom related to Minkowski space-time appear prominently in the pro\-per\-ty of covariance via the Poincar\'e group (or the dif\/feomorphism group).
\begin{itemize}\itemsep=0pt
\item
In our opinion this leaves open the possibility that a careful examination of the properties of ``modular covariance'' induced by a state $\omega$ on an algebra of observables $\As$ might reveal and identify those degrees of freedom that are related to the
``spatio-temporal structure'' of the physical system.
\item
Most of the structure of a typical f\/ield theory is already available from the simple assignment $(\As,\omega)$ without the need to resort to underlying commutative geometries:
\begin{itemize}\itemsep=0pt
\item[--]
Inside an algebra $\As$ of observables, isotony is still meaningful for the net of subal\-geb\-ras.
\item[--]
Locality can still be considered between subalgebras that are mutually commuting.
\item[--]
The specif\/ication of a state $\omega$ on the algebra immediately entails a ``local covariance'' group of GNS-implementable automorphisms of the subalgebras.
\end{itemize}
\item
As we already mentioned in Section~\ref{sec: mtp}, conditions of ``modular covariance'' (for example the ``geometric modular action'' introduced by D.~Buchholz, S.J.~Summers) have been already used to identify vacuum states and recover particular commutative space-times.

What we propose here is a kind of similar program: given a state over an algebra of observables, recover via suitable modular conditions a canonical covariant net structure and try to interpret it as a free f\/ield theory (possibly on a non-commutative space).
\item
Finally, in our specif\/ic situation, we stress that, instead of quantum f\/ield theories specif\/ied via functors from geometries to algebras,  it would be much more natural to axiomatize the existence of ``spectral functors''
$\Ss:\Ag\to\Gg$ in the ``reverse direction'' from a category~$\Ag$ of unital $C^*$-algebras, with values in a category $\Gg$ of ``geometries''. In this sense the construction of modular spectral geometries proposed at the beginning of this section could be seen as an example.
\end{itemize}

As a f\/irst step in these directions, making contact with our research program on ``categorical non-commutative geometry'' (see the companion survey~\cite{BCL6} and references therein or the talk~\cite{B2}) let us consider some of the natural categorical structures naturally emerging in our specif\/ic situation:
\begin{itemize}\itemsep=0pt
\item
Let $\omega$ be just a state (not necessarily KMS) over a unital $C^*$-algebra of observables $\Fs$.
\item
Consider the family $\Ag(\omega,\Fs)$ of all the unital $C^*$-subalgebras $\As$ of $\Fs$ such that $\omega$ becomes a faithful
KMS-state when restricted to the $C^*$-algebra $\As$.

Note that if there exists a one-parameter group $\alpha$ of automorphisms of $\As$ such that $\omega$ is $\beta$-KMS with respect to $\alpha$, such a group is necessarily unique apart from rescaling and so we can always assume $\beta=-1$ for the given state.

For all $\As\in \Ag(\omega,\Fs)$ we can construct a modular spectral geometry
$(\As_\omega,\H_\omega,\xi_\omega,K_\omega,J_\omega)$.
\item
Consider the family $\Mg(\omega,\Fs)$ of Hilbert $C^*$-bimodules $\Ms$ over pairs of algebras in $\Ag(\omega,\Fs)$ whose ``linking
$C^*$-category'' is KMS for the state $\omega$.

To every bimodule $\Ms\in\Mg(\omega,\Fs)$ over $\As_1,\As_2\in \Ag(\omega,\Fs)$ we can associate a bivariant modular spectral geometry.
\item
The family $\Mg(\omega,\Fs)$ becomes an involutive category under involution given by the dual bimodules and composition given by the tensor product of bimodules.
\item
We have a ``tautological Fell bundle'' over the $*$-category $\Mg(\omega,\Fs)$ whose total space is the disjoint union of the bimodules in $\Mg(\omega,\Fs)$ and the projection functor assigns to every $x\in \Ms$ the base point $\Ms$.
The Fell bundle is also equipped with a compatible net structure.
The (categorical) modular theory and the modular geometry of this Fell bundle and also the geometry of its ``convolution algebra'' are under investigation.
\item
Although category theory is not the main focus of this review (see the companion paper~\cite{BCL6} and the talk~\cite{B2} for more on this), we just mention that def\/initions of strict higher $C^*$-categories\footnote{Bertozzini~P., Conti~R., Lewkeeratiyutkul~W., Suthichitranont~N., Strict higher $C^*$-categories, preprint.} have been already developed and we are attempting to use them for an axiomatization of C.~Rovelli relational quantum mechanics~\cite[Section~5.6]{Ro1}\footnote{Bertozzini~P., Axiomatic formulation of Rovelli relational quantum mechanics, in preparation.}. The hope is that identifying such relational structures inside the modular net described above might shed some light on the spatial organization of the system along the lines indicated by
C.~Rovelli~\cite[Section~5.6.3]{Ro1}.
\end{itemize}

There are now several strong research projects in categorical quantum gravity (such as those proposed by J.~Baez~\cite{Ba1,Ba2}, L.~Crane~\cite{Cr1,Cr2,Cr3,Cr4}, J.~Morton~\cite{Mo} and R.~Dawe Martins~\cite{DM}), and although here we have been just touching on the (higher) categories induced by modular theory, we are very interested in the possible contacts that might develop in the near future with those more developed lines of research.

\subsection{Finding the macroscopic geometry}

As already stressed before, the Tomita--Takesaki modular f\/low becomes trivial in the classical commutative case and hence there is no hope to apply the techniques above directly to the limiting commutative case in order to f\/ind a classical geometry. Under this point of view it seems that geometry arises (via modular theory) from the interplay of the non-commutativity of the algebra of observables and the choice of a representation (specif\/ied by the state).

One, now fashionable, possibility is to try to recover a macroscopic geometry as an ``emergent residue'' from the microscopic
non-commutative geometry via some form of coarse graining.
Several techniques in these directions are being developed in dif\/ferent contexts such as:
``decoherence/einselection'' (see for example H.~Zeh~\cite{Ze} and W.~Zurek~\cite{Zu}),
``co\-he\-rent states'' (see for example B.~Hall~\cite{Hal}),
``emergence/noiseless subsystems'' (see T.~Konopka, F.~Markopoulou~\cite{KoM} and O.~Dreyer~\cite{D1,D2,D3,D4,D5,D6}),
or the ``cooling'' procedure described by A.~Connes, K.~Consani, M.~Marcolli~\cite{CCM, CM2} that, analysing the geometry of the space at dif\/ferent ``temperature scales'' $\beta$, seems particularly suited to our case.

Anyway since the obstacles in this last step seem to be not very dif\/ferent from those encountered in justifying classical mechanics as a limit of quantum mechanics, one should be aware that other very interesting roads are still open. For example, algebraic superselection theory has also been suggested as a way to obtain classical geometries out of quantum theory (see for example
K.~Fredenhagen~\cite{Fr}); and in a dif\/ferent direction, the recent works by G.~Morchio, F.~Strocchi~\cite{MS1,MS2,MS3} (see also D.~Mauro~\cite{Mau}) point to a more intriguing coexistence of classical and quantum descriptions of physics. It is for sure premature for us to make def\/inite statements in this regard.

\subsection{Connection with other approaches to quantum geometry}\label{sec: otherqg}

The most dif\/f\/icult step in a ``top-down'' approach to quantum gravity, and physics in general, is usually to make contact with ``real world phenomena'' showing how the theory might, under certain limit conditions, recover the already well-known situations.

It is reassuring to know that, in the opposite direction, several models of non-commutative space-time have been proposed in recent years via deformation and/or quantization. We are especially interested in the possible contact with quantum geometries already def\/ined in the context of loop quantum gravity such as
J.~Aastrup, J.~Grimstrup, R.~Nest semi-f\/inite spectral triples~\cite{AG1,AG2,AG3,AGN1,AGN2,AGN3,AGN4,AGNP,AGP,AGP2}.
Also the reproduction of the models of quantum space-time introduced by S.~Doplicher, K.~Fredenhagen, J.~Roberts~\cite{DFR1,DFR2,Do2,Do3,Do4} in the context of algebraic quantum f\/ield theory deserves to be investigated, being the most natural
Poincar\'e covariant non-commutative version of Minkowski space-time.

Although it is clearly premature at this stage to envisage clear connections with other already well established research programs on quantum gravity\footnote{A partial list with a classif\/ication of research programs in quantum gravity, with special attention to those related to non-commutative geometry, can be found in our survey paper~\cite[Section~5.5.1]{BCL6}.}, we feel particularly compelled to look for possible links with loop quantum gravity\footnote{See C.~Rovelli~\cite{Ro1,Ro3} and  T.~Thiemann~\cite{Th} for general treatments and references on the subject and, for example, H.~Sahlman~\cite{S}, P.~Don\'a, S.~Speziale~\cite{DS} for more elementary recent introductions.}.
First of all, since loop quantum gravity purports to be the f\/irst available example of a background independent dif\/feomorphism covariant quantum f\/ield theory, in the light of the deep structural interplay between Tomita--Takesaki theory and usual quantum f\/ield theory on a f\/ixed background, which we partially reviewed in Section~\ref{sec: mtp}, it is hardly believable that modular theory is still now virtually playing no role in the foundations of loop quantum gravity.
On the other side, although non-commutative geometry appears to be totally absent in the description of the so called ``quantum geometries'' that emerge in the loop quantum gravity kinematical description of space-time as quantum spin networks and quantum foams, thanks to the already mentioned ef\/forts by J.~Aastrup, J.~Grimstrup, M.~Paschke, R.~Nest, a class of (semi-f\/inite) spectral triples has already been introduced and is potentially ready to be used in loop quantum gravity.

Here below, as the f\/irst step in the direction of J.~Aastrup, J.~Grimstrup, M.~Paschke, R.~Nest program, following several of the suggestions mentioned in Section~\ref{sec: ac}, we start providing an alternative family of spectral triples in loop quantum gravity that arise from a direct application of C.~Antonescu, E.~Christensen's construction of spectral triples for AF $C^*$-algebras.

These new spectral triples depend on the choice of a f\/iltration of graphs and on the choice of a sequence of positive real numbers and, although very similar to J.~Aastrup, J.~Grimstrup, R.~Nest triples, they still dif\/fer from them because of the following notable points:
\begin{itemize}\itemsep=0pt
\item[--]
the Hilbert spaces on which these spectral triples naturally live are identif\/ied with subspaces of the kinematical Hilbert space of usual loop quantum gravity, without the need of resorting to tensorization with a continuous part
and a matricial part,
\item[--]
their construction does not depend on the arbitrary choices of Riemannian structures on Lie groups and of Dirac operators on the Clif\/ford bundle of such classical manifolds: their Dirac operators, coming directly from the C.~Antonescu, E.~Christensen's recipe for spectral triples for AF $C^*$-algebras, seem to be completely quantal,
\item[--]
they appear to be spectral triples and not semi-f\/inite spectral triples.
\end{itemize}

Recall that a f\/inite oriented graph $\Gamma$ (that can be either an abstract graph or a graph embedded in a three-dimensional spatial surface in a globally hyperbolic Lorentzian manifold) is given by a set of nodes $\Gamma_0$, a set of edges $\Gamma_1$ and a pair of source/target maps $s,t:\Gamma_1\to\Gamma_0$\footnote{In the formal construction it is also possible to use f\/inite oriented multigraphs with loops. Usually it is assumed that $\Hom_\Gamma(v_1,v_2):=\{e\in\Gamma_1\ | \ s(e)=v_1, t(e)=v_2\}$ consists of a unique edge if $v_1\neq v_2$ and is empty otherwise.}. By a~directed family of graphs we mean a family of graphs such that any two graphs $\Gamma^1$, $\Gamma^2$ in the family admit graph inclusions into another graph $\Gamma^3$ of the family. A family of graphs closed under ``union'' (as we suppose in the following) is clearly directed.

Let $G$ be a f\/ixed compact Lie group (that in loop quantum gravity is usually $SU(2)$).

The unconstrained kinematical Hilbert space $\H$ of loop quantum gravity is usually obtained from an inductive limit of a net
$\Gamma\mapsto \H_\Gamma$ of ``graph Hilbert spaces'' $\H_\Gamma:=L^2(G^{\Gamma_1},\mu_\Gamma)$, where
$\mu_\Gamma$ is the left invariant Haar measure on the compact Lie group $G^{\Gamma_1}$.

For any given countable chain of inclusions of f\/inite oriented graphs
$\Gamma^0\subset \Gamma^1\subset \cdots \subset \Gamma^n\subset \cdots$,
the net of $C^*$-algebras $\As_{\Gamma^1}\subset \As_{\Gamma^2}\subset \cdots\subset\As_{\Gamma^n}\subset\cdots$, generated by the holonomies associated to the respective graphs, form a f\/iltration of f\/inite dimensional $C^*$-algebras with inductive limit
$\As_{(\Gamma^n)}\subset \As$ a $C^*$-subalgebra of the algebra $\As$ generated by holonomies acting on the inductive limit Hilbert space $\H_{(\Gamma^n)}\subset \H$. The same applies to the partial observable algebras of holonomies and f\/luxes $\Fs_\Gamma$\footnote{The imposition of the Gaussian constraint amounts to factoring out in the previous Hilbert spaces the action by conjugation of $G$ on the nodes of the graph and the imposition of the dif\/feomorphism constraint consists of identifying all graphs that are mapped into each other by graph automorphisms of the inductive limit graph of the directed set of f\/inite graphs.
The overall picture remains the same: for any ``dif\/feomorphism orbit'' of a~f\/inite graph~$\Gamma$ there is a~kinematical Hilbert space~$\K_\Gamma$ that actually is the truncation of the kinematical Hilbert space~$\K$ of loop quantum gravity to the f\/inite degrees of freedom associated to the equivalence class of the graph~$\Gamma$. On $\K_\Gamma\subset \K$ the algebra of holonomies
$\As_\Gamma\subset \As$ and the algebra of partial observables $\Fs_\Gamma\subset \Fs$, i.e.~the algebra generated by holonomies and f\/luxes, continue to be represented.}.

A theorem proved by C.~Fleischhach and J.~Lewandowski, A.Oko\l\' ow, H.~Sahlmann, T.~Thiemann \cite{F, LOST} assures that in loop quantum gravity, at the purely kinematical level, the $C^*$-algebra $\Fs$ of partial observables generated by holonomies and f\/luxes admits a unique dif\/feo\-mor\-phism invariant pure state $\phi$ whose GNS-representation $(\K_\phi,\pi_\phi,\xi_\phi)$ is irreducible; this means that the von Neumann algebra $\pi_\phi(\Fs)''$ coincides with $\Bs(\K_\phi)$.

With an immediate application of C.~Antonescu, E.~Christensen's Theorem~\ref{th: ac} to the f\/il\-tra\-tion of AF $C^*$-algebras
$\Fs_{\Gamma^0}\subset\Fs_{\Gamma^1}\subset \cdots \subset \Fs_{(\Gamma^n)}$ we f\/inally obtain:
\begin{Theorem}
For every given countable inclusion $\Gamma^0\subset \Gamma^1\subset \cdots \subset\Gamma^n\subset\cdots$ of finite oriented graphs  there is a family of spectral triples $(\Fs_{(\Gamma^n)},\K_{(\Gamma^n)},D_{(\Gamma^n), (\alpha_n)})$ depending on the choice of the filtration of graphs $(\Gamma^n)$ and on a sequence of positive real numbers $(\alpha_n)$ as in Theorem~{\rm \ref{th: ac}}.
\end{Theorem}

Note that we stated the previous result for a subalgebra of observables $\Fs_{(\Gamma^n)}$, but that a~perfectly similar strategy can be applied to a subalgebra generated by holonomies $\As_{(\Gamma^n)}$.

We are still conf\/ident that an appropriate modif\/ication of this construction, dealing with the extra degrees of freedom that must be  introduced because of the presence of a Clif\/ford bundle over $G^{\Gamma_1}$ and a matricial representation $\MM_n(\CC)$ of $G$ in the def\/inition of the Hilbert spaces $L^2(G^{\Gamma_1},\mu_\Gamma)\otimes\CCl(T^*(G^{\Gamma_1}))\otimes\MM_n(\CC)$, might be adapted to match the specif\/ic case of J.~Aastrup, J.~Grimstrup, M.~Paschke, R.~Nest, but we did not have time to investigate this issue.

\medskip

Finally we would like to spend some more words on the connection of modular theory with general covariant quantum theories and, in the specif\/ic case, loop quantum gravity.

As already noted, the $C^*$-algebra $\Fs$ of partial observables is irreducibly represented on the kinematical Hilbert space $\K$ of loop quantum gravity, that we can identify with the Hilbert space~$\K_\phi$ of the GNS-representation $(\pi_\phi,\K_\phi,\xi_\phi)$ of the unique dif\/feomorphism invariant pure state $\phi$ on the algebra $\Fs$, the ``covariant Fock vacuum'', also known as the ``empty state''.

Let us, for the moment, proceed in the same direction leading from Gibbs equilibrium states (for f\/inite systems) to the KMS-condition in quantum statistical mechanics and suppose that the dynamical constraint of the theory can be specif\/ied by an operator $H$ in
$\K_\phi$, with $e^{-\beta H}$ trace-class (for $\beta\in \RR$), that is leaving f\/ixed the covariant Fock vacuum $\xi_\phi$.

On the Koopman--von Neumann Hilbert space $\K_\phi\otimes\K_\phi^*$ there is an induced one-parameter group of automorphisms generated by the symmetrized operator
$H\otimes I - I\otimes (\Lambda_\phi H \Lambda_\phi^{-1})$, where we recall that $\Lambda: \K_\phi\to\K_\phi^*$ is the usual Riesz conjugate-linear isomorphism. Applying trivially the well-known properties of modular theory for the case of the von Neumann algebra
$\B(\K_\phi)$ (see for example A.~Connes, C.~Rovelli~\cite{CR} and R.~Haag~\cite[Section~V.1.4]{H}), we see that for every inverse temperature $\beta$ there exists a unique Gibbs KMS-state with respect to the one-parameter group $t\mapsto \Ad_{e^{itH}}$ given, for $x\in \Fs$, by
$\omega_\beta(x):=\tau(\rho_\beta x)$, where $\tau$ is the canonical trace on $\B(\K_\phi)$ and
$\rho_\beta:=e^{-\beta H}/\tau(e^{-\beta H})$ is a density operator.
The $\omega_\beta$-GNS representation $(\pi_{\omega},\K_\omega,\xi_\omega)$ is in this case identif\/ied as follows:
the Hilbert space coincides with the Koopman--von Neumann's $\K_\omega:=\K_\phi\otimes\K_\phi^*$, the representation $\pi_\omega$ is just $\pi_\phi\otimes I$ i.e.~the left action of $\Fs$ on $\K_\phi\otimes\K_\phi^*$ and
$\xi_\omega:=\rho_\beta^{1/2}\in \K_\phi\otimes\K^*_\phi$ is taking the role of ``covariant vacuum''.

To the pair $(\Fs,\omega_\beta)$ we associate the modular spectral geometry
$(\Fs_\omega,\K_\omega, \xi_\omega, K_\omega,J_\omega)$, where the modular conjugation operator $J_\omega$ coincides with the conjugate-linear f\/lip operator def\/ined on simple tensors by
$J_\omega(\zeta\otimes\eta):=\Lambda_\phi^{-1}(\eta)\otimes\Lambda_\phi(\zeta)$, for all $\zeta\otimes\eta\in \K_\omega$ and the modular generator coincides, modulo the factor $\beta$, with the symmetrized constraint
$K_\omega:=\beta H\otimes I- \beta \otimes (\Lambda_\phi H\Lambda_\phi^{-1})$.
Of course the factor $\beta$ does not have ef\/fect here, since the multiplication by a constant does not modify the solution of the constraints detemined by $K_\omega$ on $\K_\omega$ and $H$ on $\K_\phi$.

By turning backwards the previous argument, we obtain the following mathematically totally trivial, but ideologically important result:
\begin{Theorem}
In loop quantum gravity, if the dynamical constraint is described $($modulo a~positive constant$)$ by an operator $H$ such that
$\tau(e^{{-\beta H}})<+\infty$ on the kinematical Hilbert space, the dynamical content of the theory is completely specified by the choice of a KMS-state on the algebra of partial observables $\Fs$.
\end{Theorem}
\begin{proof}
Just observe that, from the previous discussion $H=-\beta^{-1}(\log(\rho_\beta) -\log(\tau(e^{-\beta H}))$ where $\rho_\beta$ is the unique density operator such that $\tau(\rho_\beta x)=\omega_\beta(x)$ and reabsorb the constants.
\end{proof}

\begin{itemize}\itemsep=0pt
\item
The message that we learn from such a toy model is that, exactly as in quantum statistical mechanics (where KMS-states and modular theory subsume the theory of Gibbs equilibrium states extending it to  systems with inf\/inite degrees of freedom), in a general covariant quantum theory the dynamics can be specif\/ied by the choice of a KMS-state.
\end{itemize}

\begin{Remark}
As a general fact, without dealing with the details, we note that in this kind of toy models where there is a specif\/ic irreducible representation that is already singled out at the kinematical level (the covariant Fock vacuum), the specif\/ication of the modular spectral geometry associated to a given KMS-state $\omega$ on the algebra of partial observables $\Fs$ essentially f\/ixes the dynamical system $(\As,\H_\phi,H)$, for the a priori given algebra of ``coordinates'' $\As\subset\Fs$ via the equation
$K_\omega:=H\otimes I- \otimes (\Lambda_\phi H\Lambda_\phi^{-1})$ imposed at the level of the Koopman-von Neumann space
$\K_\omega$.
\end{Remark}

As a corollary this implies that if the algebra of partial observables $\Fs$ if an AF $C^*$-algebra, the Dirac operator obtained by C.~Antonescu, E.~Christensen construction $(\Fs,\H_\phi,D_{(\Gamma^n),(\alpha_n)})$, via the process of tensorial symmetrization described in Section~\ref{sec: ac}, can be determined (modulo a~global multiplication constant) by the specif\/ication of a KMS state
$\omega$, by imposing the equation
$D_{(\Gamma^n),(\alpha_n)}\otimes I -I\otimes(\Lambda_\phi D_{(\Gamma^n),(\alpha_n)}\Lambda_\phi^{-1})=K_\omega$ on the space $\H_\phi\otimes\H^*_\phi$. Again, at this ``toy model'' level, the specif\/ication of a KMS-states essentially determines a non-commutative geometry on the algebra of observables.

Of course much more needs to be understood in the case of more complicated $C^*$-algebras of partial observables, where such a simple ``splitting mechanism'' for the modular Hamiltonian fails, and where a full use of techniques from modular theory will be necessary: after all this is only hinting at the beginning of what might be called ``thermodynamics of covariant quantum systems'' \dots.

\subsection{Quantum physics \dots}

In this last subsection, we would like to make some extremely general considerations on the status of quantum theory as a fundamental theory and discuss the relevance of our modular quantum gravity program in this context.

Despite the realization that important conceptual problems are still af\/fecting quantum mechanics as a fundamental theory of nature (the problem of measurement, the problem of time for covariant quantum theories, the conf\/licts with classical determinism) and despite the recurrent claims of a need for a theory that supersedes, modif\/ies or extends quantum theory (hidden variables, several alternative interpretations, collapse of wave function, deterministic derivations of quantum theory), very few people have pointed out (perhaps because of a misinterpretation of N.~Bohr's correspondence principle) that quantum theory still now is essentially an incomplete theory, incapable of ``standing on its own feet''.

\textit{In all the current formulations of quantum theory, the basic degrees of freedom of a theory are specifically introduced ``by hand'' and make always reference to a classical underlying geometry.}

The choice of the degrees of freedom is usually done in elementary quantum mechanics through ``quantization procedures'' via the imposition of Weyl (or Heisenberg) commutation relations for conjugated observables starting from a classical pair of position and momentum as in Dirac canonical quantization and, more generally, as in Weyl quantization, associating quantum observables starting from functions living on a classical phase-space.

In quantum f\/ield theory of free f\/ields again, the local Weyl algebras are obtained by second quantization from symplectic spaces that originate from propagators of hyperbolic operators living on a classical Lorentzian manifold.

Even in algebraic quantum f\/ield theory, the most abstract axiomatization available, we have seen that there is always an underlying classical space-time as an indexing base of the net of quantum observables or a category of classical space-time geometries as a domain of the $C^*$-algebra-valued functor that def\/ines the theory.

What is even more incredible is that, although it is widely recognized that many of the problems encountered in quantum f\/ield theory can be traced back to ``unhealthy usage'' of classical notions of space-time manifolds (divergences, convergence of Feynman integrals), the previous problem of \textit{intrinsic characterization of degrees of freedom ``inside quantum theory''} has rarely, if ever, been seriously considered (with some notable exceptions) and that most of the attempts to cure the problem are simply trying to substitute ``by hand'' commutative classical geometries with non-commutative counterparts without addressing the issue from within quantum physics.

Among those really notable exceptions in this direction (see the companion review~\cite[Section~5.4]{BCL6} for more details), we have to mention the few attempts to spectrally reconstruct space-time from operationally def\/ined data of observables and states in algebraic quantum f\/ield theory starting from U.~Bannier~\cite{Ban} and culminating in S.J.~Summers, R.~White~\cite{SuW} and, at least form the ideological point view, the ef\/forts in information theoretical approaches for an operational def\/inition of quantum theory as, for example, in A.~Grinbaum~\cite{Gri1,Gri2,Gri3}.

A very interesting passage, asserting that in a fundamental theory the notion of space-time must be derived a posteriori and understood through relations between ``quantum events'', can be found in R.~Haag~\cite[Section~VII, Concluding remarks]{H}.

Our main ideological motivation for the work on ``modular algebraic quantum gravity'' presented here comes from the
view~\cite{B1,B3,BCL1,BCL6} that:
\begin{itemize}\itemsep=0pt
\item[]
\textit{space-time should be spectrally reconstructed a posteriori from a basic operational theory of observables and states;}
\item[]
\textit{A.~Connes' non-commutative geometry provides the natural environment where to attempt an implementation of the spectral reconstruction of space-time;}
\item[]
\textit{Tomita--Takesaki modular theory should be the main tool to achieve the previous goals, associating to operational data, spectral non-commutative geometries.}
\end{itemize}

A very interesting recent work by C.~Rovelli, F.~Vidotto~\cite[see Section~III B]{RoV} is making some clear progress in the direction of the f\/irst two claims above obtaining relations between entropy of graphs describing the quantum geometry of  loop gravity and the ``spectral geometry'' given by the non-relativistic Hamiltonian of a single particle interacting with the gravitational f\/ield.

Still the deepest and at the same time most visionary proposal for a program aiming at the reconstruction of space-time via interacting events from purely quantum theoretical constructs is the one that has been given by C.~Rovelli~\cite[Section~5.6.4]{Ro1} inside the framework of his relational theory of quantum mechanics~\cite{Ro2,RS}.

The modest ideas suggested here might be seen as just a very partial attempt to set up a~mathematical apparatus capable of implementing C.~Rovelli's intuition, merging it with non-commutative geometry and utilizing instruments from category theory,
higher $C^*$-categories and Fell bundles, in order to formulate a  theory of relational quantum mechanics.

A few comments might be necessary to provide some justif\/ication for our choice to base a~theory of quantum relativity on (a ``covariant'' and categorical variant of) the usual formalisms of \hbox{$C^*$-algebras} and states.
First of all the actual usage of $C^*$-algebras and von Neumann algebras is motivated from the technical point of view by the availability of an already sophisticated mathematical theory (notably modular theory). Suitable topological involutive algebras might be considered as well in the future. From the physical point of view, besides the classical investigations
(see~R.~Haag~\cite[Sections~I, VII]{H}), there are some strong indications that the formalism of involutive operator algebras is operationally based in a fairly general sense (see for example the notes by M.~Paschke~\cite{P3}) and furthermore, that fundamental theories expressed in terms of operator algebras can be motivated on purely logical and information theoretic grounds (see for example G.-M.~D'Ariano~\cite{D'} and references therein).

Finally we must stress that, contrary to most of the proposals for fundamental theories in physics that are usually of an ontological character, postulating basic microscopic degrees of freedom and their dynamics with the goal to explain known macroscopic behaviour, our approach (if ever successful) will only provide an absolutely general operational formalism to model information acquisition and communication/interaction between quantum observers (described via certain categories of algebras of operators) and to extract from that some  geometrical data in the form of a non-commutative geometry of the system.

Possible connections of these ideas to ``quantum information theory'' and ``quantum computation'' are also under
consideration~\cite{B1}.

\subsection*{Acknowledgements}

P.B.~wishes to thank C.~Rovelli at CPT in Marseille, J.~Barrett at the QG$^2$-2008 conference at the university of Nottingham and S.J.~Summers at the university of Florida in Gainesville, for the opportunities to discuss some of the ideas and of\/fer seminars exposing most of the original material here presented, in May, July 2008 and in April 2009.

We thank one of the anonymous referees of the paper for pointing out some missing references on modular localization and on the algebraic proof of  Bisognano--Wichmann theorem based on scattering theory in Section~\ref{sec: mtp}.

 \LastPageEnding

\end{document}